\documentclass[twocolumn,twoside]{IEEEtran}

\usepackage{mathpple}
\usepackage{times}

\usepackage{amsmath}
\usepackage{amssymb}
\usepackage{amsopn}
\usepackage{amsfonts}
\usepackage{latexsym}

\usepackage{upref}
\usepackage{theorem}

\usepackage{graphicx}
\usepackage{psfrag}
\usepackage{epsf}

\usepackage{cite}

\title{Coding for the Optical Channel:\\[-0.2ex]
the Ghost-Pulse Constraint\\[0.60ex]}

\author{\large 
Navin Kashyap,~\IEEEmembership{Member,~IEEE,}
Paul H.\ Siegel,~\IEEEmembership{Fellow,~IEEE,} 
and Alexander Vardy,~\IEEEmembership{Fellow,~IEEE}
\vspace{-3.25ex}
\thanks{Manuscript submitted September 8, 2004, revised August 18, 2005. 
This work was supported in part by the UC 
Discovery Grant Program, the~Applied Micro Circuits Corp., San Diego, CA,
the National Science Foundation, and the David and Lucile 
Packard Foundation. The work was carried out while the first author 
was at the University of California San Diego.
}
\thanks{Navin Kashyap is with the Department of Mathematics
and Statistics at Queen's University, Kingston, ON K7L 3N6, 
Canada. (email: {nkashyap@\ mast.queensu.ca}).
}
\thanks{Paul H.\ Siegel is with the Department of Electrical 
and Computer Engineering and with the Center for Magnetic 
Recording Research, University of California San Diego, 
   La Jolla, CA 92093--0407, U.S.A.
  (e-mail: {psiegel@ ucsd.edu}). 
}
\thanks{Alexander Vardy is with the 
   Department of Electrical and Computer~Engineering,
   the Department of Computer Science and Engineering,
   and the Department of Mathematics,
   all at the University of California San Diego,
   La Jolla, CA 92093--0407, U.S.A.
  (e-mail: vardy@kilimanjaro.ucsd.edu).}
}

\renewcommand{\markboth}[2]
{\renewcommand{\leftmark}{#1}\renewcommand{\rightmark}{#2}}

\theoremstyle{plain} 
\theorembodyfont{\normalfont\slshape}

\newtheorem{thm}{Theorem$\!$} 
\newenvironment{theorem}
{\begin{thm}\hspace*{-1ex}{\bf.}}{\end{thm}}

\newtheorem{lem}[thm]{Lemma$\!$}
\newenvironment{lemma}{\begin{lem}\hspace*{-1ex}{\bf.}}{\end{lem}}

\newtheorem{prop}[thm]{Proposition$\!$}
\newenvironment{proposition}{\begin{prop}\hspace*{-1ex}{\bf.}}{\end{prop}}

\newtheorem{cor}[thm]{Corollary$\!$}
\newenvironment{corollary}{\begin{cor}\hspace*{-1ex}{\bf.}}{\end{cor}}

\newtheorem{defn}{Definition$\!$}
\newenvironment{definition}{\begin{defn}\hspace*{-1ex}{\bf.}}{\end{defn}}

\newcounter{enumrom}
\renewcommand{\theenumrom}{(\roman{enumrom})}

\makeatletter
\renewcommand{\@endtheorem}{\endtrivlist}
\makeatother

\makeatletter
\renewcommand{\thefigure}{{\bf \@arabic\c@figure}}
\renewcommand{\fnum@figure}{{\bf Fig.}\,\thefigure}
\makeatother

\newcommand{\cA}{{\cal A}} 
\newcommand{\cB}{{\cal B}}

\newcommand{\cF}{{\cal F}}
\newcommand{\cG}{{\cal G}} 
\newcommand{\cH}{{\cal H}}

\newcommand{\cQ}{{\cal Q}} 

\newcommand{\cS}{{\cal S}} 
\newcommand{\cT}{{\cal T}}

\newcommand{\cX}{{\cal X}}

\newcommand{\bfit}{\bfseries\itshape}

\newcommand{\be}[1]{\begin{equation}\label{#1}}
\newcommand{\ee}{\end{equation}} 
\newcommand{\eq}[1]{(\ref{#1})}

\renewcommand{\le}{\leqslant} 
\renewcommand{\leq}{\leqslant}
\renewcommand{\ge}{\geqslant} 
\renewcommand{\geq}{\geqslant}

\newcommand{\Tref}[1]{Theo\-rem\,\ref{#1}}
\newcommand{\Pref}[1]{Pro\-po\-si\-tion\,\ref{#1}}
\newcommand{\Lref}[1]{Lem\-ma\,\ref{#1}}
\newcommand{\Cref}[1]{Coro\-lla\-ry\,\ref{#1}}
\newcommand{\Dref}[1]{De\-fin\-it\-ion\,\ref{#1}}

\newcommand{\Z}{{\mathbb Z}}

\newcommand{\deff}{\mbox{$\stackrel{\rm def}{=}$}}

\DeclareMathOperator{\supp}{supp}

\DeclareMathAlphabet{\mathbfsl}{OT1}{cmr}{bx}{it}
\newcommand{\xxx}{\mathbfsl{x}}
\newcommand{\yyy}{\mathbfsl{y}}

\newcommand{\HBGP}{H_{\rm BGP}}
\newcommand{\HTGP}{H_{\rm TGP}}

\newcommand{\Aqn}{{\cA_{q}^{n}}}

\newcommand{\al}{{\alpha}}

\newcommand{\SF}{{\cS_{\scriptscriptstyle\cF(2)}}}

\newcommand{\zero}{\mathbf{0}}
\newcommand{\one}{\mathbf{1}}

\newcommand{\xxxi}{\xxx_{\kern1pt i}}

\newcommand{\G}{{\cH}}
\newcommand{\hH}{H'}
\newcommand{\hB}{\cB'_{3;t}}
\newcommand{\hG}{\cG}
\newcommand{\hx}{x'}

\mathchardef\inn="3232
\renewcommand{\in}{\mbox{$\,\inn\,$}}

\begin{document}

\maketitle

\begin{abstract}
We consider a number of constrained coding techniques that
can be used to mitigate a nonlinear effect in the optical fiber channel
that causes the formation of spurious pulses, called ``ghost pulses.''
Specifically, if $b_1 b_2 \ldots b_{n}$ is a sequence of bits sent 
across an optical channel, such that $b_k=b_l=b_m=1$ for some $k,l,m$
(not necessarily all distinct) 
but $b_{k+l-m} = 0$, then the ghost-pulse effect causes $b_{k+l-m}$ to
change to $1$, thereby creating an error. 
Such errors do not occur if the sequence of bits satisfies 
the following constraint: 
for all integers $k,\,l,\,m$ such that $b_k = b_l = b_m = 1$, 
we have $b_{k+l-m} = 1$.
We call this the binary ghost-pulse (BGP) constraint.\
We will show, however, 
that the BGP
constraint has zero capacity, implying that sequences satisfying 
this constraint cannot carry much information. 
Consequently, we 
consider a more sophisticated coding scheme, which uses {\bfit ternary}
sequences satisfying a certain ternary ghost-pulse (TGP) constraint.
We further relax these constraints by ignoring interactions 
between symbols that are more than a certain distance $t$ 
apart in the transmitted sequence. 
Analysis of the resulting BGP$(t)$ and TGP$(t)$ constraints 
shows that these have nonzero capacities, and furthermore, 
the TGP$(t)$-constrained codes can achieve rates that are
significantly higher than those for the correponding
BGP$(t)$ codes. We also discuss the design of encoders and
decoders for coding into the BGP, BGP$(t)$  
and TGP$(t)$ constraints.
\end{abstract}

\begin{keywords}
Binary ghost-pulse\,(BGP) constraint, 
capacity of constrained systems,
constrained encoding and decoding, 
optical communication, 
ternary ghost-pulse\,(TGP) constraint.\vspace{-1ex}
\end{keywords}

\renewcommand{\thefootnote}{\arabic{footnote}}
\setcounter{footnote}{0}

\section{Introduction} 
\vspace{-.25ex}
\label{sec:Introduction}

\noindent\looseness=-1
High data-rate optical fiber communication presents several interesting 
challenges to a coding theorist. The diverse impairments 
peculiar to the optical channel necessitate the development of new coding 
schemes, capable of mitigating the effects of these impairments. 
One such impairment is the nonlinear effect known as intrachannel
four-wave mixing (FWM)--- see~\cite{essiambre}, \cite{mamyshev,shake}
and references therein. FWM results in strong inter-symbol interference  
between the symbols in a bitstream transmitted across the optical fiber.
It is widely accepted~\cite{liu,mamyshev,zweck} that at bit rates of 
40\,Gbps and beyond, FWM will play a major role in limiting the 
information-carrying capacity and the propagation distance\pagebreak[3.99] 
of a dispersion-managed optical communication system. 
In this paper, we consider a number of constrained coding techniques 
motivated by the intrachannel FWM effect.

\subsection{Background on Ghost-Pulse Formation}
\label{sec:1A}

\begin{figure}[ht]
\epsfxsize=3.00in 
\centerline{\epsfbox{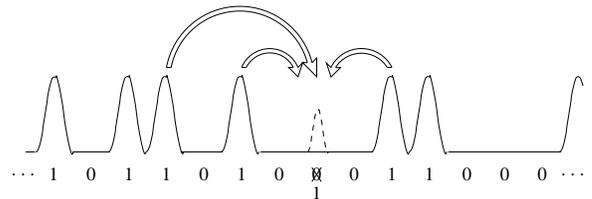}}
\vspace{-2ex}
\caption{Model of ghost-pulse formation 
due to the interaction of three pulses}
\label{ghost_pulse}
\end{figure}

\noindent\looseness=-1
In a typical optical fiber communication scenario, 
a train of light pulses, corresponding to a sequence 
of $n$ bits, is sent across an optical fiber.
Each bit in the sequence is allocated a
time slot of duration $T$, and a binary one or zero
is marked by the presence or absence of a pulse in that 
time-slot. The effect of intrachannel FWM 
is to transfer energy from triples of pulses in `1'-slots into
certain `0'-slots, thereby creating spurious pulses 
known as \emph{ghost pulses}. 
It has been observed that the interaction 
of pulses in the $k$-th, $l$-th, and $m$-th time-slots 
pumps energy into the $(k{+}l{-}m)$-th time-slot. If this slot
did not originally contain a pulse --- that is, if 
the $(k{+}l{-}m)$-th bit was a zero in the original 
$n$-bit sequence --- then this transfer of energy creates 
a ghost pulse in this time-slot.
This could cause the original zero to be read as a one 
(see Fig.\,\ref{ghost_pulse}). 

Since the overall energy is conserved, some of the pulses in 
the $k$-th, $l$-th, and $m$-th time-slots lose energy, resulting 
in a lowering of their amplitude (intensity). On the other hand, 
if the $(k{+}l{-}m)$-th slot already contained a pulse, then there 
is an exchange of energy between the pulses in the
$k$-th, $l$-th, \mbox{$m$-th}, and $(k{+}l{-}m)$-th slots, leading
to amplitude fluctuations. An analytic explanation of these
phenomena can be derived using the nonlinear Schr{\"o}dinger equation 
that describes pulse propagation in optical fibers --- see
\cite{ablowitz1,ablowitz2,zweck}.

\looseness=-1
There are, in general, multiple $(k,l,m)$ triples that result in the 
same integer $k{+}l{-}m$. Thus it is possible to have several pulse triples 
generating a ghost pulse at the same time-slot. Of course, in reality,
the number of pulse triples involved in ghost-pulse formation at 
a certain time-slot is quite small. This is because, as one would 
expect from physical considerations, the interaction between
pulses that are sufficiently far apart in the transmitted pulse train 
is weak. Indeed, for typical optical transmission parameters, pulses 
that are more 
than 10 to 12 time-slots apart do not contribute significantly to 
the formation of ghost pulses \cite{nikola,liu}. 
In any case, when multiple pulse triples generate a ghost pulse 
at the same zero time-slot, the resulting ghost pulse is the 
superposition of the ghost pulses formed by each of the pulse 
triples. The superposition of multiple ghost pulses may 
result in a stronger ghost pulse, or sometimes (due to destructive 
interference) in a weaker ghost pulse.

As shown in \cite{alic,liu}, the phases of the original pulses play 
a vital 
role in determining which pulses lose energy and which gain energy 
in the course of the energy transfer induced by FWM. 
The phase of a ghost pulse created by a given pulse triple
depends on the phases of all the pulses in the triple. Thus, in the case 
of a superposition of multiple triples, the relationship of the phase 
of the resulting ghost pulse to that of all the pulses involved in its 
creation can be quite complex. Indeed, even the amplitude of a ghost 
pulse depends on the phases of the pulse triples involved in its creation, 
since superposition of ghost pulses with opposing phases at the same 
time-slot will actually suppress ghost-pulse formation. 

Physically, a ghost pulse is just another pulse of light. 
Thus it is possible that ghost-pulse formation may propagate:
the interaction of a ghost pulse with actual pulses or other ghost 
pulses may lead to the creation of even more ghost pulses. 

Finally, it should be noted that FWM is primarily a problem with 
long-haul and ultra long-haul optical communication systems,
operating at 40\,Gbps. This is so because the amplitude of ghost 
pulses grows linearly with propagation distance. A long-haul system 
consists of many periods of alternating spans of conventional and 
dispersion-compensating fiber. This causes quasi-periodic broadening 
and compression of the~informat\-ion-bearing pulses. For typical 
transmission parameters,~ghost pulse amplitude reaches significant 
proportions over several periods of the dispersion map that is typically 
50--100\,km long. 
The simulations reported in the literature~\cite{liu,shake,zweck} 
were 
carried out over links of length 500\,km to 5000\,km.

\subsection{Related Modulation Techniques}

\noindent\looseness=-1
The optics literature has seen the emergence of several simple 
modulation schemes \cite{alic,cheng,liu,kumar} aimed at reducing the 
impact of FWM. Most of these schemes are based on the fact 
that FWM is a phase-sensitive effect and, therefore, can be
controlled by modulating the phase of the pulses being transmitted.
The one exception is the modulation scheme of~\cite{kumar}, which proposes 
to use unequally spaced pulses at the expense of sacrificing 
spectral efficiency. 

Coding --- that is, introduction of redundancy in the transmitted bits
as a means of controlling errors --- has not been given much consideration 
as an approach to mitigating the FWM effect. To the best of our knowledge,
the only previous work in this area has been reported 
by Vasic, Rao, Djordjevic, Kostuk, and Gabitov in~\cite{vasic}. In that
paper, the authors use sequences satisfying a certain maximum-transition 
run (MTR) constraint to counter the impact of FWM. In the language 
of constrained coding, a binary sequence $\xxx$ is said to satisfy 
an MTR($j$) constraint 
if every run of ones in $\xxx$ has length at most~$j$ (cf.\,\cite{moon}).
In the modulation scheme of~\cite{vasic}, a block code of rate 0.8, 
consisting of $256$ binary codewords of length 10 satisfying the 
MTR$(2)$ constraint, is used for transmission. 
Simulation results show significant ghost-pulse reduction 
due to the use of this coding scheme. The authors of~\cite{vasic} conclude that 
``it is possible to successfully tackle the detrimental effects 
of FWM in 40-Gb/s systems using simple 
coding techniques.''

In this paper, we undertake a systematic study of a number of 
coding schemes 
that combine constrained coding and phase modulation. 
Our study focuses purely on the coding-theoretic aspects 
(e.g.\ rate, encoding/decoding) of these schemes --- we make
no claims regarding their effectiveness in suppressing ghost pulses. 
In particular, we do not address the question of how well 
constrained coding schemes are suited to tackle the problem 
of eliminating ghost pulses in real-world optical systems. 
Such questions can only be answered via experimentation and/or
extensive simulations of the fiber-optic channel, which
is beyond the scope of this work.

\subsection{Binary Ghost-Pulse Constraint}

\noindent\looseness=-1
To formulate a well-defined coding problem, we model the formation of 
(primary) ghost pulses 
as follows. Let $b_1 b_2 \ldots b_{n}$, 
with $b_i \in \{0,1\}$, 
be the binary sequence corresponding to the
train of pulses sent across the fiber optic medium.
If for some integers $k$, $l$, and $m$ (not necessarily all distinct),
we have 
\be{BGP-def}
b_k = b_l = b_m = 1
\hspace{3ex}\text{while}\hspace{2ex}
b_{k+l-m} = 0
\ee
then the formation of a primary ghost pulse converts the zero
in time-slot $k{+}l{-}m$ to a one.  
Note that if we can encode the transmitted binary sequence
in such a way that~\eq{BGP-def} never occurs, we will
eliminate all (higher-order) ghost pulses caused by ghost-pulse
propagation (discussed in Section\,\ref{sec:1A}), as well.
For example, a sequence containing at most one $1$, or 
the all-ones sequence, or a sequence of alternating zeros 
and ones all satisfy this condition.
In general, we say that a binary sequence 
$c_1 c_2 \ldots c_{n}$ satisfies the 
\emph{binary ghost-pulse (BGP) constraint}
if for all integers $k,l,m$ such that $c_k=c_l=c_m = 1$
and $0 \leq k\,{+}\,l\,{-}\,m \leq n{-}1$, 
we also have $c_{k+l-m} = 1$. It is clear that 
transmitting a sequence that satisfies the
BGP constraint will not allow ghost pulses to be created.

Let $f_{\rm BGP}(n)$ be the number of binary sequences of 
length~$n$ that satisfy the BGP constraint. Then
the asymptotic information rate (or the \emph{capacity}, 
or the \emph{entropy}) of the BGP constraint is defined
(cf.~\cite{LM,HCT}) as follows: 
\be{HBGP_def}
\HBGP 
\ \deff\ 
\lim_{n \to \infty} \frac{1}{n} \log_2 f_{\rm BGP}(n). 
\ee
Of course, we would like $\HBGP$ to be as close to $1$
as possible, so that coding into the BGP constraint
adds little redundancy to the information being encoded. 
However, as we will show in Section~\ref{bgp_section}, a finite-length
binary sequence satisfies the BGP constraint if and only if the 
ones in the sequence are uniformly spaced --- that is, the positions 
of the ones form an arithmetic progression. 
It follows that there are $O(n^2)$ binary sequences 
of length $n$ that satisfy the BGP constraint, and 
$\HBGP = 0$.
Hence, we need to investigate alternative approaches 
to dealing with the ghost-pulse problem.

\looseness=-1
One approach that we consider 
is based on the intuition that the interaction between 
pulses that are sufficiently far apart in the transmitted 
pulse train is weak. As noted in Section\,\ref{sec:1A},
in a typical optical communication scenario, pulses that are 
more than 10--12 time-slots apart do not contribute significantly 
to the formation of ghost pulses (cf.~\cite{nikola,liu}). 

Disregarding the interaction between ones that are 
separated by more than some fixed distance $t$, we say
that a binary sequence $c_1 c_2 \ldots c_{n}$ satisfies
\emph{the BGP$(t)$ constraint} if for all integers $k,l,m$
(not necessarily distinct) such that
\begin{eqnarray}
&&c_k = c_l = c_m = 1,
\\[.75ex]
&&
0 \,\leq\, k\,{+}\,l\,{-}\,m \,\leq\, n-1, \hspace*{14ex}\\[-4.5ex] \nonumber
\end{eqnarray}
and\vspace{-.50ex}
\be{max-t}
\max\bigl\{|k\,{-}\,l|,|l\,{-}\,m|,|m\,{-}\,k|\bigr\} \,\leq\, t 
\ee
we also have $c_{k+l-m} = 1$.
The capacity $H_2(t)$ of the BGP$(t)$ constraint can be
defined as in 
\eq{HBGP_def}. (We will provide formal definitions 
for the capacities of all such constraints 
in the next section.) 
In Section\,\ref{bgp_section}, we will show that $H_2(t)$ 
is positive for all $t$. 
However, we also show in Section\,\ref{bgp_section}
that $H_2(t)$ lies in the range $0.21$\,--\,$0.25$, 
when $t \in \{10,11,12\}$.
This makes the BGP$(t)$
constraint somewhat unattractive as the basis for 
a~coding scheme. Nevertheless, we 
briefly discuss in Section\,\ref{bgp_section} the design 
of finite-state encoders that take an unconstrained binary 
sequence as input and produce a BGP$(t)$-constrained sequence 
as output.

\subsection{Ternary Ghost-Pulse Constraint}
\label{sec:1D}

\noindent\looseness=-1
Another approach that has been suggested~\cite{alic,cheng,liu} to 
mitigate the formation of ghost pulses 
is to apply, at the transmitter end, a phase shift of 
$\pi$ to some of the pulses. 
We can effectively think of this phase-modulation technique 
as converting a binary sequence $b_1 b_2 \ldots b_{n}$,
with $b_i \in \{0,1\}$, into a ternary sequence
$c_1 c_2 \ldots c_{n}$, where $c_i \in \{-1,0,1\}$, such that
$b_i = |c_i|$ for all $i$.
One reason behind this phase-modulation approach is that, 
as explained in Section\,\ref{sec:1A}, superposition of the 
contributions due to multiple pulse triples will result in 
suppression of ghost-pulse formation if their interference is
destructive. Thus, knowledge of the relationship between the 
phase of a ghost pulse and the phases of the pulses involved 
in its creation makes it possible to manipulate the phase of 
the transmitted pulses in a way that encourages destructive
interference. Such phase modulation schemes are very effective 
at eliminating some of the stronger ghost pulses (cf.~\cite{alic,liu}). 
However, as observed in \cite{alic}, it is impossible to achieve 
destructive interference in several consecutive zero-slots.
Moreover, these schemes do not mitigate the ``side ghosts'' that 
arise due to energy leakage from the one-slots into adjacent 
zero-slots. Therefore, another approach is to modulate the phase 
of the transmitted pulses with the aim of achieving energy 
redistribution among the one-slots, thereby preventing energy 
leakage into adjacent zero-slots. Overall, building upon the 
work of~\cite{alic}, it appears reasonable to try preventing situations 
in which pulses in time-slots $k$, $l$, and
$m$ all have the same phase, while the slot at time $k{+}l{-}m$ 
is empty\,(zero).

\looseness=-1
Thus we say that a ternary sequence $c_1 c_2 \ldots c_{n}$ 
satisfies \emph{the ternary ghost-pulse (TGP) constraint}
if for all integers 
$k$,~$l$,~$m$ 
(not necessarily distinct) 
such that $0 \leq k\,{+}\,l\,{-}\,m \leq n{-}1$,~and 
\be{plusminus}
c_k = c_l = c_m = +1
\hspace{3ex}\text{or}\hspace{3ex}
c_k = c_l = c_m = -1
\ee
we also have $c_{k+l-m} \neq 0$. Let $\cT_3$ be 
the set of all finite-length ternary sequences 
that satisfy the TGP constraint. To transmit 
a finite-length binary data sequence, we\pagebreak[3.99]
encode it as a sequence from $\cT_3$. Based on 
the discussion above, we shall assume, as a first-order
approximation, that sequences in $\cT_3$ are effective 
in mitigating ghost-pulse formation, so the
transmitted sequence can be recovered without error 
at the receiver end.

However, there is a catch. Most long-haul optical 
communication systems 
use direct-detection optical receivers, 
which can only detect the intensity (amplitude) 
of the optical signal at the channel output, 
not its phase. Thus if the transmitted ternary sequence
was $c_1 c_2 \ldots c_{n}$, then the receiver only sees the sequence
$|c_1|, |c_2|, \ldots, |c_{n}|$. In other words,
the receiver cannot
distinguish a $+1$ from a $-1$. As a result, we cannot use two sequences
in $\cT_3$ that differ only in phase (sign) to encode
two different binary data sequences. 

\looseness=-1
We thus have a rather unusual coding problem:
even though the sequence being transmitted is ternary,
the alphabet used for encoding information is effectively binary. 
In general, discrete channels for which 
the output alphabet is \emph{smaller} than 
the input alphabet
are rarely encountered in information theory. In fact, to the 
best of our knowledge, a situation where the alphabet over which
the constraint is defined is different from the information-bearing
alphabet has not been previously studied in the constrained 
coding literature.

In order to describe the procedure for encoding a binary 
data sequence using TGP-constrained sequences, we define the set 
\be{B3-def}
\cB_3 
\ \deff\,\
\bigl\{
|c_1|, |c_2|, \ldots, |c_{n}| ~:~ c_1 c_2 \ldots c_{n} \in \cT_3
\bigr\}.
\ee
This is the set of all finite-length \emph{binary} sequences that can 
be converted to a sequence in $\cT_3$ by changing certain $1$'s to $-1$'s.
To transmit a binary data sequence $a_1 a_2 \ldots a_{N}$, we 
first encode it as a sequence $b_1 b_2 \ldots b_{n} \in \cB_3$, 
which is 
then converted to a corresponding sequence $c_1 c_2 \ldots c_{n} \in \cT_3$ 
at the input to an optical channel. At the channel output, the receiver 
detects the sequence $b_1 b_2 \ldots b_{n}$, which can be 
uniquely decoded to recover the original binary sequence 
$a_1 a_2 \ldots a_{N}$. 

The capacity $\HTGP$ of the TGP constraint can be 
now defined in a manner analogous to \eq{HBGP_def}. 
Let $f_{\rm TGP}(n)$ denote the number of sequences of length $n$
in the set $\cB_3$. Then
\be{HTGP_def}
\HTGP 
\ \deff\ 
\lim_{n \to \infty} \frac{1}{n} \log_2 f_{\rm TGP}(n).
\ee
The analysis of the TGP constraint appears to be a much more
difficult problem 
than analysis of the BGP constraint. However, we conjecture that 
$
\HTGP = \HBGP = 0
$.
Strong evidence in support of this conjecture is given in~\cite{KSV} (see 
Section\,\ref{tgp_section}).

Consequently, we consider the weaker TGP$(t)$ constraint obtained,
similarly to the BGP$(t)$ constraint, by ignoring interactions 
between nonzero symbols that are more than distance $t$ apart. 
Define the set $\cT_{3;t}$ by adjoining the extra condition \eq{max-t}
to \eq{plusminus}. The capacity $H_3(t)$ of the TGP$(t)$ constraint
can be then defined as in \eq{HTGP_def}, but with respect 
to the set $\cT_{3;t}$ rather than $\cT_3$.
One can reasonably expect that as $t$ increases, $H_3(t)$ decreases, 
converging upon $\HTGP$ in the limit as $t \to \infty$.
Indeed, we will {prove} in the next section that 
\be{HTGP-claim}
\HTGP \, = \, \lim_{t \to \infty} H_3(t) \, = \ \inf_{t \geq 1} H_3(t).
\ee
This provides a means of computing increasingly tight upper bounds 
on the capacity $\HTGP$ which, as we mentioned earlier, is not easy to compute
directly. Furthermore, we will show in Section\,\ref{sec:4}
that 
\be{H3-results}
H_3(1) \, = \, 1
\text{~~and~~}
H_3(2) \, \approx \, 0.96.
\ee
These values are significantly larger than the corresponding values
for the BGP$(t)$ constraint, namely
\be{H2-results}
H_2(1) \, \approx \, 0.69
\text{~~and~~}
H_2(2) \, \approx \, 0.55.
\ee
Moreover, it appears from \eq{H3-results} and \eq{H2-results}
that $H_3(t)$ decreases much slower with $t$ than $H_2(t)$, since 
$H_2(1)-H_2(2) \approx 0.14$ 
while 
$H_3(1)-H_3(2) \approx 0.04$.
Assuming that this trend continues for larger values of $t$, 
coding schemes based on TGP$(t)$-constrained sequences 
can be a reasonably efficient means of mitigating the 
ghost-pulse effect in optical communications. 

Unfortunately,
the techniques we use in Section\,\ref{sec:4}
to determine $H_3(1)$ and $H_3(2)$ do 
not easily generalize to the computation of $H_3(t)$ for arbitrary $t$.
Thus we have been unable to verify whether the aforementioned trend continues 
for larger values of $t$. 
In Section\,\ref{tgpt_section}, we describe 
a general method for computing $H_3(t)$; however,
this method is too computationally intensive to be implemented in practice. 
Nevertheless, we do discuss (also in Section\,\ref{sec:4})
the design of finite-state encoders for coding schemes involving 
TGP$(t)$-constrained sequences.\vspace{1ex}

\noindent
{\bf Remark.}
Before concluding this introductory section, we note that it is possible
to design other coding schemes that combine constrained coding 
with phase modulation in order to achieve ghost-pulse suppression. 
For example, we can conceivably add
phase modulation to the constrained coding 
scheme of \cite{vasic}, thereby gaining some improvement in performance. 
In this paper, however, we have chosen to focus solely on the BGP and 
TGP constraints. The unusual nature of these constraints requires the 
development of non-standard tools for their anal\-ysis, which 
may be of independent interest to coding theorists.

\section{Definitions and Preliminary Results}
\label{def_section}

\noindent
In this section, we formally define the various types of ghost-pulse
constraints that we shall be interested in.
We also give precise definitions for the corresponding
capacities, and establish several useful relationships 
between them.

Let $\Z$ and $\Z^+$ denote the set of integers and the set
of positive integers, respectively. Given $n, n' \in \Z$, 
we write
\begin{eqnarray*}
[n] 
& \hspace{-1ex}\deff\hspace{-1ex} &
\bigl\{\,i \in \Z ~:~ 1 \leq i \leq n\,\bigr\}
\\[-0.25ex]
[n,n'] 
& \hspace{-1ex}\deff\hspace{-1ex} &
\bigl\{\,i \in \Z ~:~ n \leq i \leq n'\,\bigr\}.
\end{eqnarray*}
Note that both $[n]$ and $[n,n']$ could be empty.
Let $\cA_2 = \{0,1\}$ and let $\cA_3 = \{-1,0,1\}$.
These are the relevant alphabets for the binary and  
the ternary ghost-pulse constraints, respectively.
However, rather than giving definitions for the 
binary case and the ternary case separately, we
find it more convenient to define the ghost-pulse 
constraints over a generic $q$-ary alphabet. 
Thus, given an integer $q \ge 2$, let $\cA_q$ denote
a fixed set of $q$ letters, one of which is a distinguished
letter $0$. 
Although this is not required in what follows,
a good way to think of $\cA_q$ is as the set 
of distinct $q-1$th roots of unity, augmented by zero. 
For $n \in \Z^+$, let $\Aqn$ denote 
the set of sequences of length $n$ over $\cA_q$.
Given $\xxx = (x_1 x_2 \ldots x_n) \in \Aqn$, the
\emph{support of $\xxx$} is defined as 
$
\supp(\xxx) 
=  
\{i \in [n] : x_i \neq 0\}
$.

\begin{definition}
\label{qgp_def}
A sequence $\xxx \in \Aqn$ satisfies the $q$-ary ghost-pulse ($q$-GP)
constraint if for all $k,l,m \in\! \supp(\xxx)$ 
such that 
$$
x_k \,=\, x_l \,=\, x_m
$$
either\, $k+l-m \in \!\supp(\xxx)$ or\, $k+l-m \,{\notin}\, [n]$. Note 
that the integers $k,l,m$ above 
are not necessarily distinct.
\end{definition}

For $n \in \Z^+\!$, let $\cT_q(n)$ be the set of 
sequences of length $n$ over $\cA_q$ that satisfy 
the $q$-GP constraint. 
Further define 
\be{Tq-def}
\cT_q \ \deff\, \bigcup_{n = 1}^{\infty} \cT_q(n).
\ee
This is the set of all \emph{finite-length} 
sequences satisfying the $q$-GP constraint. 
Let $\xi: \cA_q \to \cA_2$ be the ``absolute value'' function,
defined by
\be{xi-def}
\xi(x)
\ \deff \
\left\{
\begin{array}{@{\hspace{0.5ex}}rl}
0 & x = 0 \\
1 & x \ne 0.
\end{array}
\right.
\ee
For all $n \in \Z^+\!$,
we extend this ``absolute value'' function componentwise to a function 
$\xi: \Aqn \to \cA_2^n$ via 
\be{xi-vector-def}
\xi(x_1 x_2 \ldots x_{n})
\ \deff \
\bigl(\xi(x_1), \xi(x_2), \ldots, \xi(x_{n})\bigl).
\ee
Given such a function, we further define for all $n \in \Z^+\!$ the sets 
$\cB_q(n) \subset \cA_2^n$ as follows
\be{Bq-def}
\cB_q(n) 
\, = \,
\xi\bigl(\cT_q(n)\bigr) 
\,\ \deff \,\ 
\bigl\{\,\xi(\xxx) \,:\, \xxx \in \cT_q(n) \,\bigr\}.
\ee
Finally, we set $\cB_q = \xi(\cT_q) = \bigcup_{n=1}^{\infty} \cB_q(n)$. 
Thus, if $\cA_q{\setminus}\{0\}$
is indeed a set of complex roots 
of unity, then $\cB_q(n)$, respectively $\cB_q$, consists of those 
\emph{binary} sequences that can be transformed into a sequence 
in $\cT_q(n)$, respectively $\cT_q$, by means of appropriate
phase shifts. In particular, our definition of $\cB_3$ based upon
\eq{Bq-def} coincides with the earlier definition in \eq{B3-def}.

\begin{definition}
\label{Hq-def}
For all integers $q \ge 2$, the capacity of the 
$q$-ary ghost-pulse constraint is defined by
\be{Hq_def}
H_q 
\,\ \deff\ \lim_{n \to \infty} \frac{\log_2 |\cB_q(n)|}{n}.
\end{equation}\vspace{-1.50ex}
\end{definition}

It should be immediately clear from the discussion above that 
$H_2 = \HBGP$ and $H_3 = \HTGP$, as defined in \eq{HBGP_def} 
and \eq{HTGP_def} respectively. The following proposition shows 
that all these capacities are, indeed, well-defined.

\begin{proposition}
\label{limit-prop}
The limit below exists for all $q \ge 2$, and moreover
\be{Hq_inf_eqn}
\lim_{n \to \infty} \frac{\log_2|\cB_q(n)|}{n}
\ = \
\inf_{n \geq 1} \frac{\log_2|\cB_q(n)|}{n}.\vspace{1ex}
\ee
\end{proposition}

\begin{proof}
This follows from the standard argument for shift spaces 
(see e.g.\ \cite[pp.\,103--104]{LM}), which we briefly reproduce
here for completeness. Use the following test for convergence
from elementary calculus: if $a_1,a_2,\ldots$ is a sequence of 
nonnegative numbers such that $a_{n+n'} \leq a_n + a_{n'}$ for 
all $n,n' \geq 1$, then $\lim_{n \to \infty} a_n/n$ exists and 
equals $\inf_{n \geq 1} a_n/n$. 
Apply this test to the sequence defined by $a_n = \log_2 |\cB_q(n)|$.
We need to show that $a_{n+n'} \leq a_n + a_{n'}$ or, equivalently,
that
\be{Prop1-aux}
\left|\cB_q(n\,{+}\,n')\right| 
\, \leq \,
\left|\cB_q(n)\right|
\left|\cB_q(n')\right|.
\ee
But this easily follows from the observation 
that if a sequence $\yyy \in \cA_q^{n+n'}$ satisfies the $q$-GP
constraint, then every contiguous subsequence of $\yyy$ also 
satisfies the $q$-GP constraint. Hence if 
$(x_1 x_1 \ldots x_{n+n'}) \in \cB_q(n{+}n')$, 
then 
$(x_1 x_2 \ldots x_{n}) \in \cB_q(n)$ 
and
$(x_{n+1} x_{n+2} \ldots x_{n+n'}) \in \cB_q(n')$,
which implies \eq{Prop1-aux}.
\vspace{1.5ex}
\end{proof}

\looseness=-1
We will show in Section\,\ref{sec:3A} 
that $H_2 = 0$, and our analysis in Section\,\ref{tgp_section}
will lead us to conjecture that $H_3 = 0$ as well.
In fact, we believe that $H_q = 0$ for all $q$. This
is so because the $q$-GP constraint has unbounded 
memory. For large $n$, the value of a sequence $\xxx \in \cT_q(n)$
at a given position $i \in [n]$ depends on the values of 
$\xxx$ at (essentially) {all} other positions. 
To obtain nonzero capacities, we relax the $q$-GP
constraint by bounding its effective memory, 
as made precise in the next definition. As explained 
in Section\,\ref{sec:Introduction}, 
it makes physical sense to do so.

\begin{definition}
\label{qgpt_def}
Let $t \in \Z^+\!$ be fixed. 
A sequence $\xxx \in \Aqn$ satisfies the \mbox{$q$-GP$(t)$}
constraint if for all 
$k,l,m \in\! \supp(\xxx)$ 
such that
$$
x_k \,=\, x_l \,=\, x_m
\hspace{2.5ex}\text{\rm and}\hspace{2.5ex}
\max\bigl\{|k{-}l|,|l{-}m|,|m{-}k|\bigr\} \,\leq\, t
$$
either\, $k+l-m \in \!\supp(\xxx)$ or\, $k+l-m \,{\notin}\, [n]$. 
As before, the integers $k,l,m$ above 
need not be all 
distinct.\vspace{0.50ex}
\end{definition}

For $n,t \in \Z^+$, we let $\cT_{q;t}(n)$ denote the set of 
sequences of length $n$ over $\cA_q$ satisfying the $q$-GP$(t)$
constraint, and define
$
\cT_{q;t} = \bigcup_{n=1}^{\infty} \cT_{q;t}(n)
$
as in~\eq{Tq-def}.
With the help of the function $\xi: \Aqn \to \cA_2^n$ 
given by \eq{xi-def} and \eq{xi-vector-def}, we define
\be{Bqt-def}
\cB_{q;t}(n) 
\, = \,
\xi\bigl(\cT_{q;t}(n)\bigr) 
\,\ \deff \,\ 
\bigl\{\,\xi(\xxx) \,:\, \xxx \in \cT_{q;t}(n) \,\bigr\}
\ee
and write $\cB_{q;t} = \xi(\cT_{q;t}) = \bigcup_{n=1}^{\infty} \cB_{q;t}(n)$. 
We can now define the capacity of the $q$-GP$(t)$
constraint 
as follows.

\begin{definition}
\label{Hqt-def}
For all integers $q \ge 2$ and $t \ge 1$, the capacity of the $q$-GP$(t)$
constraint is defined by
\be{Hqt_def}
H_q(t) 
\,\ \deff\ \lim_{n \to \infty} \frac{\log_2 |\cB_{q;t}(n)|}{n}.
\ee\vspace{-2.0ex}
\end{definition}

Exactly the same argument that we used in the proof of~\Pref{limit-prop}
can be now used to show that the limit in \eq{Hqt_def} exists,
and in fact
\be{Hqt_inf_eqn}
H_q(t)
\, = \,
\inf_{n \geq 1} \frac{\log_2 |\cB_{q;t}(n)|}{n}.
\ee
Observe that, for all fixed $q$, 
the sequence $H_q(1),H_q(2),\ldots$
is a nonincreasing 
sequence of nonnegative numbers. 
This is so because 
$\cB_{q;t+1}(n) \subseteq \cB_{q;t}(n)$ for all $n \in \Z^+$
and all $t \in \Z^+$,
as is evident from \Dref{qgpt_def}.
Therefore $\lim_{t \to \infty} H_q(t)$
exists, and equals $\inf_{t \geq 1} H_q(t)$. 
The following proposition shows that this 
limit is also equal to $H_q$, as defined
in \eq{Hq_def}.  

\begin{proposition}
\label{Hq_Hq*}
For all integers $q \ge 2$, 
\be{Prop2}
H_q
\, = \,
\lim_{t \to \infty} H_q(t) 
\, = \,
\inf_{t \geq 1} H_q(t).\vspace{-.5ex}
\ee
\end{proposition}

\begin{proof}
Let $\al_q \,\deff\, \inf_{t \geq 1} H_q(t)$. We have 
already shown that $\lim_{t \to \infty} H_q(t) = \al_q$,
so it remains to prove that $H_q = \al_q$. It 
follows immediately from \Dref{qgp_def} and \Dref{qgpt_def}\pagebreak[3.99]
that $\cB_{q;t}(n) \supseteq \cB_q(n)$ for all $n,t \in \Z^+$.
Hence
$|\cB_{q;t}(n)| \geq |\cB_q(n)|$
and $H_q(t) \geq H_q$ for all $t \in \Z^+$. 
Letting $t \to \infty$, we conclude 
that $\al_q \geq H_q$.
For the reverse inequality, first fix an $n \in \Z^+$
and observe that $\cB_q(n) = \cB_{q;n}(n)$. Therefore
$$
\frac{\log_2 |\cB_q(n)|}{n} 
\, = \,
\frac{\log_2 |\cB_{q;n}(n)|}{n} 
\ \geq\  
\inf_{m \geq 1} \frac{\log_2 |\cB_{q;n}(m)|}{m}.
$$
Note that the right-hand side above is precisely $H_q(n)$
in view of~\eq{Hqt_inf_eqn}, and $H_q(n) \ge \al_q$
by the definition of $\al_q$. If follows that 
$\log_2|\cB_q(n)|/{n} \ge \al_q$ for all $n \in \Z^+\!$,
and therefore $H_q \geq \al_q$. This completes
the proof of the proposition.  
\vspace{1.0ex}
\end{proof}

Observe that our claim in \eq{HTGP-claim} 
follows as a special case (for $q=3$) from \Pref{Hq_Hq*}.
Thus, as discussed in Section\,\ref{sec:1D}, 
\Pref{Hq_Hq*} provides a means of computing increasingly 
tight upper bounds on $H_q$. In particular, this proposition 
implies that $\HTGP = H_3$ can be 
determined by studying the asymptotics of the
sequence $H_3(1), H_3(2),\ldots$. In Section~\ref{tgpt_section}, we
show that there is indeed an algorithm that can be used to 
compute $H_3(t)$ for any given $t$. 
Unfortunately, this algorithm is too computationally 
intensive to be useful in practice.

\section{The Binary Ghost-Pulse Constraints}
\vspace{.250ex}
\label{bgp_section}

\noindent
Following the terminology of Section\,I,
we shall refer to the $q$-GP constraints with $q=2$ 
as the \emph{binary ghost-pulse} (BGP) constraints.
Such constraints can be completely analyzed, and 
the purpose of this section is to present this analysis.

\subsection{The BGP Constraint with Unbounded Memory}
\label{sec:3A}

\noindent
Note that Definitions~\ref{qgp_def} and \ref{qgpt_def}
become somewhat redundant in the binary case.
For a binary sequence $\xxx = (x_1 x_2 \ldots x_{n})$,~\emph{any}  
$k,l,m \in \supp(\xxx)$ satisfy \mbox{$x_k = x_l = x_m$}. 
Thus the BGP constraint is simply the requirement that for all 
$k,l,m \in\kern-1pt \supp(\xxx)$, either $k+l-m \in\kern-1.5pt \supp(\xxx)$ 
or $k+l-m {\notin} [n]$. 
The following theorem makes use of this observation to show 
that sequences that satisfy the BGP constraint are 
precisely those whose support set forms an arithmetic
progression.

\begin{theorem}
\label{S2n_theorem}
For all $n \in \Z^+\!$, a sequence $\xxx \,{=}\, (x_1 x_2 \ldots x_{n}) \in\cA_2^n$ 
satisfies the BGP 
constraint 
iff there exist $a,d \in [0,n]$ such that 
\be{th3}
\supp(\xxx) \, = \, (a+d\Z) \cap [n].
\ee
\end{theorem}

\begin{proof}
$(\Leftarrow)$ Suppose that $\xxx$ satisfies \eq{th3}, 
and consider any $k_1,k_2,k_3 \in \supp(\xxx)$. Then
$k_i = a+ d j_i$ for some $j_i \in \Z$. 
Set $j = j_1 {+} j_2 {-} j_3$.
Then $k_1 + k_2 - k_3 = a+jd \in \kern-1pt (a+d\Z)$. Thus,
either $k_1+k_2-k_3 \in\kern-1.5pt \supp(\xxx)$ 
or $k_1+k_2-k_3 {\notin} [n]$. 

$(\Rightarrow)$ Suppose that $\xxx  = (x_1 x_1 \ldots x_{n})$ 
satisfies the BGP constraint.
If $\supp(\xxx) = \varnothing$, then 
we can take $a = d = 0$ in \eq{th3}.
If $|\supp(\xxx)| = 1$, then we can take
$a$ to be the unique integer in $\supp(\xxx)$ and set $d=0$. 
Hence, it remains to consider the case where $|\supp(\xxx)| \geq 2$. 
For this case, set
\be{d_def}
d 
\, = \, 
\min \bigl\{\, |k - m| ~:~  k,m \in\! \supp(\xxx),\ k \neq m\,
\bigr\},
\ee
and then take $a$ to be any integer with 
$a, a+d \in \supp(\xxx)$. 
To prove that $\xxx$ satisfies \eq{th3} with this choice of 
$a$ and $d$, we will first 
show that $(a+d\Z) \cap [n] \subset \supp(\xxx)$, and then 
prove that every element of $\supp(\xxx)$ must also be 
in $a + d\Z$.

{\bf Claim\,1:} $(a+d\Z) \cap [n] \subset \supp(\xxx)$.
In order to establish this claim, suppose that 
\be{th3-claim1}
\bigl\{\,a+\ell d ~:~ i \le \ell \le j
\,\bigr\} \, \subset\, \supp(\xxx)
\ee
for some $i \leq 0$ and $j \geq 1$. By our 
choice of $a$ and $d$, we know that \eq{th3-claim1}
certainly holds for $i = 0$ and $j = 1$. Observe that 
\begin{eqnarray*}
a + (i{-}1)d  
& \hspace{-1ex}=\hspace{-1.5ex} &
\bigl(a+id\bigr) + \bigl(a+id\bigr) - \bigl(a+(i{+}1)d\bigr)
\end{eqnarray*}
and
\begin{eqnarray*}
a + (j{+}1) d 
& \hspace{-1ex}=\hspace{-1.5ex} &
\bigl(a+jd\bigr) + (a+jd) - \bigl(a + (j{-}1)d\bigr).
\end{eqnarray*}
Hence, if $\xxx$ satisfies the BGP constraint, then
$a+(i{-}1)d$ and $a+(j{+}1)d$ belong to $\supp(\xxx)$,
provided only that these positions are in $[n]$.
In other words, we can grow the arithmetic progression
on the left-hand side of \eq{th3-claim1} in both directions,
as long as it fits inside $[n]$, and the claim follows.
\vspace{1ex}

\begin{figure}[ht]
\epsfxsize = 3.30in
\centerline{\epsfbox{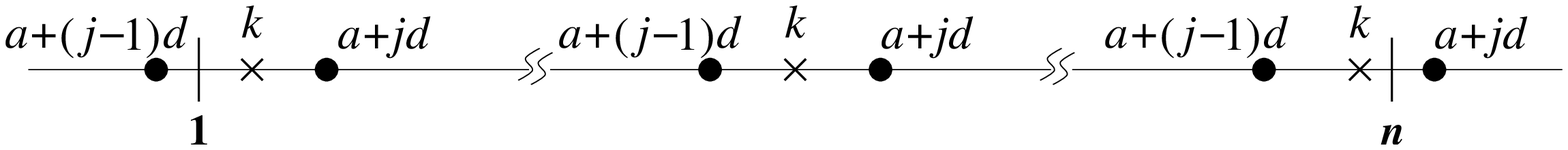}}
\caption{\hspace*{-1.0ex}Various possibilities for the choice 
of $k \in \kern-1.5pt\supp(\xxx)$ with $k \,{\not\inn}\, (a+d\Z)$}
\label{bgp_fig}
\end{figure}

{\bf Claim\,2:} $\supp(\xxx) \,{\subset}\, (a + d\Z)$. Assume to the 
contrary that there is a $k \in \supp(\xxx)$ with 
$k {\notin} (a+d\Z)$. Then we must have
\be{th3-claim2}
a+(j{-}1)d \ < \ k \ < \ a+jd
\ee
for some $j \in \Z$ such that at least one of 
$a+(j{-}1)d$ and $a+jd$ lies in $[n]$ (cf. Fig.\,2). 
Without loss of generality (w.l.o.g.),~suppose that 
$(a + jd) \in [n]$. Then $(a + jd) \in \supp(x)$ 
in view of Claim\,1. But the difference between 
$a + jd$ and $k$ is strictly less than $d$ by \eq{th3-claim2},
which contradicts the definition of $d$~in~\eq{d_def}.\vspace{1ex}

\noindent
By Claim\,1 and Claim\,2, we have 
$\supp(\xxx) = (a+d\Z) \cap [n]$,
which completes the proof of the theorem.\vspace{1.0ex}
\end{proof}

\begin{corollary}
\label{H2_cor}
There are at most\, $(n+1)^2$ sequences in $\cA_2^n$ 
that satisfy the BGP constraint, and therefore
$$ 
\HBGP 
\,\ \deff\ 
\lim_{n \to \infty} \frac{\log_2\! |\cB_2(n)|}{n}
\, \le \,
\lim_{n \to \infty} \frac{\log_2 (n\,{+}\,1)^2}{n}
\, = \, 
0.
$$
\end{corollary}

\begin{proof}
There are $(n+1)^2$ different ways of selecting the integers 
$a$ and $d$ from $[0,n]$. By \Tref{S2n_theorem},
every sequence in $\cB_2(n)$ is uniquely determined by
one such choice.\vspace{1ex}
\end{proof}

In fact, using \Tref{S2n_theorem} as a starting point,
a more careful analysis of the possible choices for $a$ and $d$
shows that 
\be{B2-enumeration}
|\cB_2(n)| 
\ = \,
\left\{\hspace{-.5ex}
\begin{array}{l@{\hspace{4.5ex}}l}
\frac{1}{4}(n+2)(n+2) & \text{$n$ even}
\\[1.0ex]
\frac{1}{4}(n+1)(n+3) & \text{$n$ odd}.
\end{array}
\right.
\ee
We leave the proof of this expression as a straightforward, 
but tedious, combinatorial exercise for the reader.

\subsection{The BGP$(t)$ Constraints}
\label{sec:3B}

\noindent
We next take on the analysis of the BGP$(t)$ constraint, 
for arbitrary $t \in \Z^+$. We will show that the BGP$(t)$
constraint is closely related to the well-known 
{$(t,\infty)$ 
constraint}. A binary sequence $\xxx$ 
is said to \emph{satisfy the $(t,\infty)$ constraint} if there
are at least $t$ zeros between any two ones in $\xxx$.
We use $\cS_{t,\infty}\kern-1pt(n)$ to denote the set of all 
$(t,\infty)$-constrained binary sequences of length $n$.
Such sequences have been extensively studied in the
constrained coding literature~\cite{Immink,Immink2,ISW,LM,HCT}.
The~next theorem shows that the set $\cB_{2;t}(n)$ of
all sequences in $\cA_2^n$ that satisfy the 
BGP$(t)$ constraint   
is not much larger than
$\cS_{t,\infty}\kern-1pt(n)$.

\begin{theorem}
\label{S2nt_theorem}
Let $\cQ_t(n)$ denote the set of all 
sequences $\xxx \in \cA_2^n$
such that\, $\supp(\xxx) = (a+d\Z) \cap [n]$ 
for some $a$ and $d$ in $[0,t]$. 
Then for all $n,t \in \Z^+$, we have
\be{th5}
\cB_{2;t}(n) 
\ = \
\cS_{t,\infty}\kern-1pt(n) \cup \cQ_t(n).
\ee
\end{theorem}

\begin{proof}
It is easy to see from (the proof of) \Tref{S2n_theorem}
that $\cQ_t(n) \subset \cB_{2;t}(n)$. Note that if 
$\xxx \in \cS_{t,\infty}\kern-1pt(n)$, then 
\eq{max-t} cannot be satisfied by any 
$k,l,m \in \supp(\xxx)$. Hence, by \Dref{qgpt_def}, all
$\xxx \in \cS_{t,\infty}\kern-1pt(n)$ also belong to 
$\cB_{2;t}(n)$. It follows that
\be{inc1}
\Bigl(\cS_{t,\infty}\kern-1pt(n) \cup \cQ_t(n)\Bigr)
\, \subseteq \,
\cB_{2;t}(n). 
\ee
To establish the inclusion in the other direction, it 
would suffice to show that 
\be{inc2}
\Bigl(\cB_{2;t}(n) \setminus \cS_{t,\infty}\kern-1pt(n)\Bigr)
\, \subseteq\,
\cQ_t(n).
\ee
Thus consider an 
$\xxx \in\kern-1pt \bigl(\cB_{2;t}(n)\kern1pt
{\setminus}\kern1pt \cS_{t,\infty}\kern-1pt(n)\bigr)$. 
Since $\xxx {\notin} \cS_{t,\infty}\kern-1pt(n)$, there exist 
distinct $k,m \in \supp(\xxx)$ with $|k-m| \leq t$. Define
\be{defd-aux}
d 
\ \, \deff\ \, 
\min \bigl\{\, |k - m| ~:~  k,m \in\! \supp(\xxx),\ k \neq m\,
\bigr\}
\ee
as in \eq{d_def}, and note that $1 \leq d \leq t$. As in \Tref{S2n_theorem},
let $a'$ be any integer with $a', a'+d \in \supp(\xxx)$. 
Then exactly the same argument we used in the proof of 
\Tref{S2n_theorem} shows that 
\be{supp-aux}
\supp(\xxx) = (a'+d\Z) \cap [n].
\ee
Finally, set $a = a' \!\!\mod{d}$.
Since $d \le t$ in \eq{defd-aux}, we obviously have $a \in [0,t{-}1]$. 
But $a'+d\Z = a+d\Z$, so \eq{supp-aux} implies that
$
\supp(\xxx) = (a+d\Z) \cap [n]
$.
Thus $\xxx \in \cQ_t(n)$, as desired. 
\vspace{1.5ex}
\end{proof}

Let $C(t,\infty)$ denote the capacity of the $(t,\infty)$
constraint,~gi-ven by
$
C(t,\infty) = 
\lim_{n \to \infty}\kern-1pt \log_2\! |\cS_{t,\infty}\kern-1pt(n)|/n
$.
It is well known
(see e.g.~\cite[p.\,88]{Immink})
that $C(t,\infty) = \log_2 \rho_t$, where $\rho_t$ is the 
largest-magnitude root of the polynomial $z^{t+1} - z^t - 1$.
It is also known that this root is always real, irrational~\cite{AS},
and lies in the open interval $(1,2)$. Thus $0 < C(t,\infty) < 1$.

\begin{table}[tp]
\centering
\caption{Capacity of the BGP$(t)$ constraint 
for $t = 1,2,\ldots,20$\vspace{-.50ex}}
\label{H2t_table}
\begin{tabular}{||cc||cc||cc||cc||}
\hline\hline
& & & & & & & \\[-1.50ex]
$t$ & $H_2(t)$ & $t$ & $H_2(t)$ &
$t$ & $H_2(t)$ & $t$ & $H_2(t)$ \\[0.50ex]
\hline
& & & & & & & \\[-1.50ex]
1  &  0.6942 & 6  &  0.3282 & 11 &  0.2301 & 16 &  0.1813 \\
2  &  0.5515 & 7  &  0.3011 & 12 &  0.2180 & 17 &  0.1742 \\
3  &  0.4650 & 8  &  0.2788 & 13 &  0.2073 & 18 &  0.1678 \\
4  &  0.4057 & 9  &  0.2600 & 14 &  0.1977 & 19 &  0.1618 \\
5  &  0.3620 & 10 &  0.2440 & 15 &  0.1891 & 20 &  0.1564 \\
\hline\hline
\end{tabular}
\end{table}

\begin{corollary}
\label{H2t_cor}
Let $\rho_t$ denote the largest-magnitude 
root of the polynomial\, $z^{t+1} - z^t - 1$.
Then for all $t \in \Z^+\!$, the capacity of the 
BGP\/$(t)$ constraint is given by
\be{cor6}
H_2(t) \,=\, C(t,\infty) \,=\, \log_2\rho_t.
\ee
\end{corollary}

\begin{proof}
This follows immediately from \Tref{S2nt_theorem}.
By~\eq{th5}, we have
$
|\cS_{t,\infty}\kern-.5pt(n)| 
\leq 
|\cB_{2;t}(n)| 
\leq 
|\cS_{t,\infty}\kern-.5pt(n)| + |\cQ_t(n)|
$. 
Note that $|\cQ_t(n)| \leq (t{+}1)^2$, as there
are $(t{+}1)^2$ different ways of choosing $a,d \in [0,t]$. The corollary
now follows from \eq{Hqt_def}.
\vspace{1ex}
\end{proof}

\looseness=-1
It is well known~\cite[p.\,89]{Immink} 
(and obvious) that $\rho_t$ decreases 
as $t$ increases. Moreover 
$\lim_{t \to \infty} \log_2\rho_t = 0$, which by Lemma\,\ref{Hq_Hq*}
provides an independent confirmation of \Cref{H2_cor}.

For reference, we list in Table\,\ref{H2t_table} the value of 
$H_2\kern-1pt(t) \,{=} \log_2\kern-.5pt\rho_t$, 
rounded to four decimal places, for all 
$t = 1,2,\ldots,20$. As can be seen from this
table, $H_2(t)$ is less than $0.25$ for all $t \ge 10$.
This means that codes consisting of sequences that
satisfy the BGP or the BGP$(t)$ constraints are not particularly
{efficient} means of mitigating the ghost-pulse problem.

\vspace{0.25ex}
\subsection{Coding Into the BGP Constraints}
\label{sec:3C}

\noindent
Nevertheless, it may still be of interest to suggest methods for encoding
an arbitrary binary sequence into a sequence satisfying the BGP or the
BGP$(t)$ constraints. 

For the BGP constraint, \Tref{S2n_theorem} and \eq{B2-enumeration}
give a precise enumeration of all the sequences in $\cB_2(n)$. Thus
unconstrained binary data can be mapped into BGP-constrained sequences 
using an enumerative coding technique~\cite{cover}.

In principle, enumerative coding can be also used to
code into the BGP$(t)$ constraints. 
However, this requires precise enumeration of the
sequences in $\cB_{2;t}(n)$ for each $n \in \Z^+$.~~Unfortunately, 
\Tref{S2nt_theorem} does not yield 
a simple formula for computing $|\cB_{2;t}(n)|$ 
as a function of $n$ and $t$. Thus, enumerative coding
would be unnecessarily complex in this case. 

\looseness=-1
We can code into the BGP$(t)$ constraint with significantly 
lower complexity if we are willing to suffer a marginal
loss in coding rate. When $n$ is sufficiently large, we can
ignore the contribution of $\cQ_t(n)$ to 
$\cB_{2;t}(n)$
for all practical purposes. Observe that when $t$ is fixed, 
$|\cQ_t(n)|$ is bounded by the constant $(t{+}1)^2$ while
$|\cS_{t,\infty}(n)|$ grows exponentially with $n$.
\vspace{0.75ex}

\begin{figure}[ht]
\epsfxsize = 2.5in
\centerline{\epsfbox{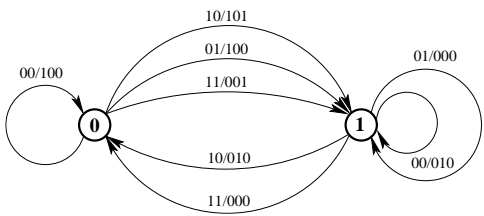}\vspace*{0.50ex}}
\caption{\hspace*{-1.0ex}A rate 2:3 
sliding-block decodable
encoder for the $(1,\infty)$ constraint}
\label{1inf_fig}
\end{figure}

Coding into the $(t,\infty)$ constraint is a very 
well-studied subject~\cite{Immink2,ISW},\cite[Chapter\,5]{LM},\cite{HCT}.
For all positive integers $p$ and $q$ with  $p/q < C(t,\infty)$,
there is a rate $p\,{:}\,q$ finite-state encoder for
the $(t,\infty)$ constraint, meaning a finite-state
machine that generates an output block of $q$ bits 
for every input block of $p$ bits, and converts
unconstrained binary sequences into sequences that satisfy
the $(t,\infty)$ constraint. For example, the graph
in Fig.\,3 is a rate 2:3 two-state 
encoder for the $(1,\infty)$ constraint.
Such rate $p\,{:}\,q$ encoders can, in fact, 
be designed so that the constrained sequences 
they generate are amenable to decoding with a 
\emph{sliding-block decoder} \cite[Theorem\,3.35]{HCT}.
For example, the encoder in Fig.\,3 is indeed sliding-block decodable: 
a description of the corresponding sliding-block decoder 
can be derived from~\cite[Example\,5.5.5]{LM}.

It is well known~\cite{AS} that the capacity $C(t,\infty)$ 
is \emph{irrational} for all $t \ge 1$. Thus the design 
and the implementation of rate $p\,{:}\,q$ encoders necessarily
becomes more cumbersome as the rate $p/q$ approaches capacity.
Consequently, in situations where variable-rate encoding and
state-dependent decoding are acceptable, the constrained coding
technique of~\cite{bender_wolf,lee}, known as ``bit-stuffing,''
is an attractive alternative.
The bit-stuffing encoder comprises two components. 
The first is an invertible distribution transformer 
that converts a sequence of i.i.d. equiprobable information bits 
into a sequence of i.i.d.\ biased bits, with the probability of 
a zero given by a prescribed value $p$.  The second component inserts
(stuffs) a string of $t$ consecutive zeros following every one 
in this biased sequence. The decoder simply discards the string 
of $t$ zeros that follows each one, and then applies 
the inverse of the distribution transformer. It can be 
shown~\cite{bender_wolf} that, if the parameter $p$ is optimized, 
the average rate of the bit-stuffing encoder equals the 
capacity $C(t,\infty)$.

\section{The Ternary Ghost-Pulse Constraints}
\label{sec:4}

\noindent
It happens to be much harder to analyze the TGP 
and TGP$(t)$ constraints than their binary counterparts
BGP and BGP$(t)$.
Nevertheless, we will attempt to do so in this section.

\subsection{The TGP Constraint with Unbounded Memory}
\label{tgp_section}

\noindent\looseness=-1
In order to gain some understanding of the structure 
of finite-length TGP-constrained binary sequences, 
we extend the definition of the TGP constraint 
in a natural way to \emph{bi-infinite 
sequences} --- that is, sequences indexed by the set of integers $\Z$.

\begin{definition}
\label{binfinite_def}
A bi-infinite sequence $\xxx \,{=}\, \smash{\{x_j\}_{j \inn \Z}}$ over 
the ternary alphabet $\cA_3 \,{=}\, \{-1,0,1\}$ 
is said to satisfy the TGP constraint if 
for all $k,l,m \in \Z$ 
such that $x_k$, $x_l$, and $x_m$ are equal and nonzero, we also 
have $x_{k+l-m} \neq 0$.
\end{definition}

Let $\cT_3^*$ denote the set of all bi-infinite
ternary sequences satisfying the TGP constraint, 
and let $\cB_3^* = \xi\bigl(\cT_3^*\bigr)$ denote the set of 
all \emph{binary} bi-infinite sequences that can be converted to
a sequence in $\cT_3^*$ by 
changing some of their $1$'s to $-1$'s. 
Using results from a branch of mathematics known as Ramsey 
theory\cite{graham}, we have shown in \cite{KSV} that any 
$\yyy \in \cB_3^*$ is \emph{almost periodic}:
it differs from a periodic sequence in 
at most two positions. 
Based on this and other results, we conjecture 
that the capacity $\HTGP = H_3$ of the TGP constraint
is zero. 

\begin{table}[ht]
\centering
\caption{Values of $\cB_3(n)$ 
for $n = 1,2,\ldots,32$\vspace{-1.50ex}}
\label{tgp_table}
\begin{tabular}{||cc||cc||cc||cc||}
\hline\hline
& & & & & & & \\[-1.50ex]
$t$ & $|\cB_3(n)|$ & $t$ & $|\cB_3(n)|$ &
$t$ & $|\cB_3(n)|$ & $t$ & $|\cB_3(n)|$ \\[0.50ex]
\hline
& & & & & & & \\[-1.50ex]
1  &     2 &   9 &   240 &  17 &  2591 &  25 & 11497 \\[0.25ex]
2  &     4 &  10 &   358 &  18 &  3245 &  26 & 13427 \\[0.25ex]
3  &     8 &  11 &   501 &  19 &  3977 &  27 & 15521 \\[0.25ex]
4  &    16 &  12 &   705 &  20 &  4881 &  28 & 17952 \\[0.25ex]
5  &    32 &  13 &   937 &  21 &  5850 &  29 & 20498 \\[0.25ex]
6  &    60 &  14 &  1248 &  22 &  7026 &  30 & 23449 \\[0.25ex]
7  &   100 &  15 &  1609 &  23 &  8313 &  31 & 26590 \\[0.25ex]
8  &   162 &  16 &  2078 &  24 &  9860 &  32 & 30193 \\[0.25ex]
\hline\hline
\end{tabular}
\end{table}

In Table\,\ref{tgp_table}, we list the number of sequences in 
$\cB_3(n)$ for all $n\,{=}\,1,2,\ldots,32$. 
All the values in Table\,\ref{tgp_table} have been 
found by exhaustive 
computer search. We then used these values  
to plot $\log_2|\cB_3(n)|/n$ as a function of $n$ in 
Fig.\,4. As can be seen from this plot, the value of
$\log_2|\cB_3(n)|/n$ decreases steadily as $n$ increases, lending
some further credence to our conjecture that 
$\HTGP = \lim_{n \to \infty} \log_2|\cB_3(n)|/n = 0$.

\begin{figure}[tb]
\epsfxsize = 2.95in
\centerline{\epsfbox{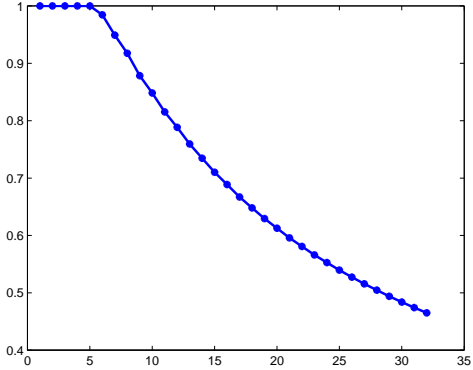}}
\caption{\hspace*{-1.0ex}Plot of $\log_2|\cB_3(n)|/n$ as a function of $n$,
for $n\,{=}\,1,2,\ldots,32$}
\label{tgp_fig}
\end{figure}

\subsection{The TGP\/$(1)$ Constraint}
\label{tgp1_section}

\noindent
For the degenerate case $t=1$, things remain simple.
It is easy to show that the set $\cB_{3;1}(n)$
of all the binary sequences that satisfy the TGP$(1)$
constraint is, in fact, the entire space $\cA_2^n$. 
This is based upon the following simple observation.
A ternary sequence $(x_1 x_2 \ldots x_{n})$ 
is in $\cT_{3;1}(n)$ if and only if the following holds:
for all $k \in [n]$ such that 
\be{TGP1}
x_k \,=\, x_{k+1} \,=\, +1
\hspace{4ex}\text{\rm or}\hspace{4ex}
x_k \,=\, x_{k+1} \,=\, -1
\ee
we have $x_{k-1} \,{\neq}\, 0$ if $(k{-}1) \in [n]$
and $x_{k+2} \,{\neq}\, 0$ if $(k{+}2) \in [n]$.
On the other hand,
it is easy to allocate signs to any binary sequence
in such a way that \eq{TGP1} never holds. In what
follows, we will often use $+$ and $-$ to denote $+1$ and $-1$,
respectively.\vspace{1ex}

\begin{figure}[ht]
\epsfxsize = 2.5in
\centerline{\epsfbox{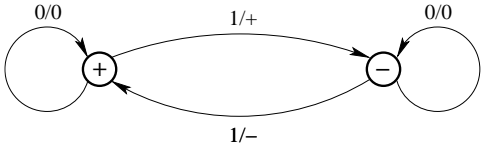}\vspace*{0.50ex}}
\caption{\hspace*{-1.0ex}Simple rate 1:1 two-state 
encoder for the TGP$(1)$ constraint}
\label{TGP(1)_fig}
\end{figure}

\begin{theorem}
\label{S3n1_theorem}
For all $n \in \Z^+\!$, we have $\cB_{3;1}(n) \,{=}\, \cA_2^n$
and therefore the capacity of the TGP\/$(1)$ constraint
is $H_3(1) = 1$.
\end{theorem}

\begin{proof}
Given any sequence $\yyy \in \cA_2^n$, the following 
encoding rule converts $\yyy$ to a ternary sequence $\xxx$
satisfying the TGP$(1)$ constraint: 
label the ones in $\yyy$ with alternating signs. More
precisely, if we think of $\yyy$ as the input to
the rate 1:1 encoder in Fig.\,5, then $\xxx$ is 
the output of the encoder. To see that $\xxx$ indeed
satisfies the TGP$(1)$ constraint, note that the
alternating signs rule guarantees that \eq{TGP1} never occurs.
\vspace{1.5ex}
\end{proof}

Observe that, in addition to its use in the proof of \Tref{S3n1_theorem},
the encoder of Fig.\,5 gives a 
practical method by which
an arbitrary finite-length binary sequence can be transformed 
into a ternary sequence satisfying the TGP$(1)$ constraint.

\subsection{The TGP\/$(2)$ Constraint}
\label{tgp2_section}

\noindent
For $t=2$, things become much more interesting. Our main result 
for this case 
is the characterization of the set 
$\cB_{3;2}{=}\kern1.5pt \xi(\kern-.5pt\cT_{3;2}\kern-1pt)$ 
of all finite-length binary sequences that satisfy the TGP$(2)$ 
constraint in terms of a small number of \emph{forbidden blocks}.

To make this 
precise, let us first clarify our use 
of the term {sub-block}. We say that a sequence
$(x'_1 x'_2 \ldots x'_{m})$ is \emph{a sub-block of the sequence} 
$(x_1 x_2 \ldots x_{n})$ if there exists an $i \in [0,n{-}m]$ 
such that
$
(x'_1 x'_2 \ldots x'_{m})
=
(x_{i+1} x_{i+2} \ldots x_{i+m})
$.
Now, let
\be{F(2)-def}
\cF(2) 
\,\ \deff \,\
\bigl\{(011100),\,(001110),\,(001111100)\bigr\}
\ee
and let $\SF(n)$ be the set of all binary sequences of length $n$
that do not contain any element of $\cF(2)$ as a sub-block. Our
main result in this subsection is the following theorem.

\begin{theorem}
\label{S3n2_theorem}
\,For all $n \in \Z^+$, we have
\be{th8}
\cB_{3;2}(n) \, = \, \SF(n).
\vspace{1ex}
\ee
\end{theorem}

We split the proof of \eq{th8} 
into two lemmas: one shows that $\SF(n) \subseteq \cB_{3;2}(n)$, 
the other establishes 
$\cB_{3;2}(n) \subseteq \SF(n)$. 
One of the two directions is easy, as the next lemma shows.

\begin{lemma}
\label{S3n2_necessity_lemma}
\,For all $n \in \Z^+$, we have
\be{lem9}
\cB_{3;2}(n) \,\subseteq\, \SF(n).
\ee
\end{lemma}

\begin{proof}
We need to show that none of the sequences in $\cB_{3;2}$
contains any of the three sequences in $\cF(2)$ as a sub-block.
Consider first the sequence $(011100) \in \cF(2)$. The three
ones in $(011100)$ can be labeled in $2^3$ different ways by $+/-$ to 
produce ternary sequences. However, noting that 
a ternary sequence $\xxx$ satisfies the TGP$(2)$
constraint if and only if so does the sequence $-\xxx$, it is
enough to consider the following four labelings of $(011100)$:
\be{011100}
\begin{array}{c@{\hspace{2ex}}l}
(0{+}{+}{+}00), & (0{+}{+}{-}00) \\
(0{+}{-}{+}00), & (0{+}{-}{-}00).
\end{array}
\ee
It can be verified by direct inspection that none of the 
four sequences in \eq{011100} satisfies the TGP$(2)$ constraint.
Hence, none can be a sub-block of a sequence in $\cT_{3;2}$, which
implies that $(011100)$ cannot be a sub-block of a sequence in $\cB_{3;2}$.
The other two forbidden blocks in $\cF(2)$ can be disposed of 
in the same way. 
\vspace{1ex}
\end{proof}

To establish inclusion in the opposite direction, we describe
an encoding rule that takes an arbitrary sequence $\yyy \in \SF(n)$
and assigns a $+/-$ labeling to the ones in $\yyy$ in such a way
that the resulting ternary sequence satisfies the TGP$(2)$ constraint.
More precisely, we construct a function 
\be{Psi}
\Psi :~
\smash{\bigcup}_{\hspace*{-2.75ex}\raisebox{-1.50ex}{$\scriptstyle n=1$}}^%
{\hspace*{-2.0ex}\raisebox{1.50ex}{$\scriptstyle \infty$}}
\!\smash{\cA_2^n} 
\,\to\,
\smash{\bigcup}_{\hspace*{-2.75ex}\raisebox{-1.50ex}{$\scriptstyle n=1$}}^%
{\hspace*{-2.0ex}\raisebox{1.50ex}{$\scriptstyle \infty$}}
\!\smash{\cA_3^n} 
\ee
such that $\xi\bigl(\Psi(\yyy)\bigr) \,{=}\, \yyy$ for all $\yyy$ 
in the domain of $\Psi$ and, furthermore, 
$\Psi(\yyy) \in \cT_{3;2}(n)$ for all $\yyy \in \SF(n)$. 
This function $\Psi$ will be 
based upon the alternating signs idea of \Tref{S3n1_theorem};
however, a much more careful analysis is now required.
 
The first step in the construction of $\Psi$ 
consists of decomposing a binary sequence $\yyy$ 
into its maximal runs. Henceforth, we use
$\zero^{j}$ and $\one^{j}$ to denote the all-zero
and the all-one sequences of length $j$, respectively. 
Any finite-length nonzero binary sequence $\yyy$
can be written uniquely in its maximal-run form:
\be{runs_in_y}
\yyy
\ = \
\bigl(\zero^{a_0} \,\one^{b_1} \,\zero^{a_1} \,\one^{b_2} \,\zero^{a_2} \,
\cdots \,\zero^{a_{r-1}}\one^{b_r} \,\zero^{a_r}\bigr)
\ee
for some $r \,{\geq}\, 1$, where $a_1,a_2,\ldots,a_{r-1}$
and\, $b_1,b_2,\ldots,b_r$ are positive integers while
$a_0,a_r \geq 0$.
Each of the $r$ sub-blocks $\one^{b_i}$ of $\yyy$ 
is called a \emph{maximal run of ones} in $\yyy$.

The next step is to convert maximal runs
into sequences over the alphabet $\{+,-\}$.
Specifically, we define the function
$\psi: 
\bigcup_{j = 1}^{\infty} \{\one^{j}\} 
\to
\bigcup_{j = 1}^{\infty} \{+,-\}^{j}$ 
as follows:
\be{psi}
\hspace*{-.650ex}\begin{array}{c}
\psi(\one^1) \,=\, {+}\,, \hspace{2ex}
\psi(\one^2) \,=\, {+}{-}\,, \hspace{2ex}
\psi(\one^3) \,=\, {+}{-}{+}
\\[1.5ex] 
\psi(\one^4) \,=\, {+}{-}{-}{+}\,, \hspace{3ex}
\psi(\one^5) \,=\, {+}{-}{+}{+}{-}
\\[1.5ex]
\psi(\one^6) \,=\, {+}{-}{-}{+}{+}{-}\,, \hspace{2.5ex}
\psi(\one^{j}) = {+}{-}{-}\,\one^{j-6}{-}{-}{+}\hspace{-1ex}
\end{array}
\ee
where the last expression above applies for all $j \ge 7$.
Observe that $\psi(\one^j)$ is a sequence of length $j$,
so that 
$\xi\bigl(\psi(\one^j)\bigr) = \one^j$.
More importantly, $\psi(\one^j)$ satisfies the 
property described in the following lemma.

\begin{lemma}
\label{psi-lemma}
Let $j$ be a positive integer other than $3$ or $5$,
and let\,
$
\psi(\one^j) 
= 
(x_1 x_2 \ldots x_{j})
$.
Then for all $k,l,m \in [j]$ such that 
\be{lem10}
\hspace*{-1ex}x_k \,=\, x_l \,=\, x_m
\hspace{1.25ex}\text{\rm and}\hspace{1.0ex}
\max\bigl\{|k{-}l|,|l{-}m|,|m{-}k|\bigr\} \leq 2
\ee
we have $k\,{+}\,l\,{-}\,m \in [j]$ as well. Thus $x_{k+l-m} \ne 0$ and,
moreover, if $(x_1 x_2 \ldots x_{j})$ is a sub-block of a 
TGP\/$(2)$-constrained ternary sequence $\xxx$
of length~$n$, then this sub-block does not impose constraints on any
of the other $n-j$ positions in $\xxx$.
\end{lemma}

\begin{proof}
The fact that $k\,{+}\,l\,{-}\,m \in [j]$ whenever
\eq{lem10} is satisfied follows by direct inspection
from \eq{psi}.
\vspace{1ex}
\end{proof}
 
Now let $\yyy \in \cA_2^n$ 
be an arbitrary binary sequence of length~$n$.
If $\yyy = \zero^n$ or $y = \one^n$, we simply set $\Psi(\yyy) = \yyy$. 
Otherwise, we decompose $\yyy$ into its maximal runs as in \eq{runs_in_y},
and set 
\be{runs_in_x}
\Psi(\yyy)
\, = \,
\bigl(\zero^{a_0}\,\xxx_1\,\zero^{a_1}\,\xxx_2\,\zero^{a_2}\, 
\cdots \ \zero^{a_{r-1}}\xxx_r\,\zero^{a_r}\bigr)
\ee
where $\xxx_i \in \{+,-\}^{b_i}$ are 
defined by the following iterative procedure:
\begin{eqnarray} 
\label{x1}
\hspace*{-2ex}\xxx_1
&\hspace{-1ex}=\hspace{-1ex}& 
\left\{\begin{array}{cl} 
{+}{+}{-}        & \text{if $a_0 = 0$ and $b_1 = 3$} \\[0.50ex]
\psi(\one^{b_1}) & {\rm otherwise}
\end{array}
\right.
\\[1.50ex]
\label{x_i}
\hspace*{-2ex}\xxxi
&\hspace{-1ex}=\hspace{-1ex}& 
\left\{\begin{array}{c@{\hspace{3.5ex}}l} 
\psi(\one^{b_i})  & 
\text{if last symbol of $\xxx_{i-1}$ is $-$}\\[0.50ex]
-\psi(\one^{b_i}) & \text{if last symbol of $\xxx_{i-1}$ is $+$}
\end{array}
\right.
\\[0.1ex] \nonumber
\end{eqnarray} 
\noindent
for all $i = 2,3,\ldots,r$, but with two exceptions.
If $b_i = 5$ while $a_{i} \in \{0,1\}$, we modify the expression for $\xxxi$
as follows:
\begin{eqnarray} 
\label{xi-mod}
\hspace*{-3ex}\xxxi
&\hspace{-1ex}=\hspace{-1ex}& 
\left\{\begin{array}{c@{\hspace{3.5ex}}l} 
{+}{-}{-}{+}{-} & \text{if last symbol of $\xxx_{i-1}$ is $-$}\\[0.50ex]
{-}{+}{+}{-}{+} & \text{if last symbol of $\xxx_{i-1}$ is $+$}\; .
\end{array}
\right.
\end{eqnarray} 
Finally, if $\yyy$ ends with $0111$ (that is, if $a_r = 0$ and $b_r = 3$),
then we also modify the expression for $\xxx_r$ as follows:
\begin{eqnarray} 
\label{xr}
\hspace*{-3ex}\xxx_r
&\hspace{-1ex}=\hspace{-1ex}& 
\left\{\begin{array}{c@{\hspace{3.5ex}}l} 
{+}{-}{-} & \text{if last symbol of $\xxx_{r-1}$ is $-$}\\[0.50ex]
{-}{+}{+} & \text{if last symbol of $\xxx_{r-1}$ is $+$}\;.
\end{array}
\right.
\end{eqnarray} 
Observe that \eq{x1}\,--\,\eq{xr} iteratively determine
$\xxx_1,\xxx_2,\ldots,\xxx_r$ in such a way that
the first symbol of $\xxxi$ is always opposite
in sign to the last symbol of $\xxx_{i-1}$, 
for all $i = 2,3,\ldots,r$. This is
the appropriate generalization of the \emph{alternating signs} 
rule of \Tref{S3n1_theorem} for the case of the TGP$(2)$ constraint. 

\begin{lemma}
\label{S3n2_sufficiency_lemma}
The function $\Psi$ defined by equations {\rm \eq{runs_in_y}\,--\,\eq{xr}}
has the following properties:\vspace{0.50ex}
\begin{quote}
\begin{itemize}
\item[\bf P1.]
For all $\yyy \in \cA_2^n$, we have 
$\xi\bigl(\Psi(\yyy)\bigr) \,{=}\, \yyy$.
\vspace{0.50ex}
\item[\hspace*{3ex}\bf P2.]
For all $\yyy \in \SF(n)$, we have 
$\Psi(\yyy) \in \cT_{3;2}(n)$.
\end{itemize}
\end{quote}
\end{lemma}

\begin{proof}
Property {\bf P1} 
means that 
$\Psi\kern-1.00pt$ converts a\kern-0.25pt\ given 
binary~sequence $\yyy$ to a ternary sequence solely
by assigning $+/-$ labes to the ones in $\yyy$. This
should be obvious from the fact that 
$\xi\bigl(\psi(\one^j)\bigr) = \one^j$
and our construction of $\Psi$ in 
\eq{runs_in_x}\,--\,\eq{xr}.

\looseness=-1
To establish property {\bf P2}, consider an
arbitrary $\yyy \in \SF(n)$ and let 
$\xxx = (x_1 x_2 \ldots x_{n})$ denote its image $\Psi(\yyy)$
under $\Psi$. We need to show that $\xxx$ satisfies the TGP(2) 
constraint. Clearly, if $\yyy \in \{\zero^n,\one^n\}$, 
then $\xxx = \yyy$ trivially satisfies the constraint. We 
therefore assume that $\yyy \notin \{\zero^n,\one^n\}$,
which implies that $\xxx$ is given by \eq{runs_in_x}. 
Now, let $k,l,m \in \supp(\xxx)$ and suppose that
$$
\hspace*{-1ex}x_k \,=\, x_l \,=\, x_m
\hspace{2.25ex}\text{\rm and}\hspace{2.0ex}
\max\bigl\{|k{-}l|,|l{-}m|,|m{-}k|\bigr\} \leq 2.
$$
We will further assume w.l.o.g.\ that $k \le l \le m$.
Clearly, either $x_k$ and $x_m$ come from the same sub-block
$\xxxi$ of $\xxx$ in \eq{runs_in_x}, or they belong to distinct 
sub-blocks $\xxxi$ and $\xxx_j$. 
This leads to two cases, which we consider next.

{\bf Case\,1:} 
$x_k$ belongs to $\xxxi$ while $x_m$ belongs to $\xxx_j$, 
with $i \ne j$.\\
Since distinct sub-blocks in \eq{runs_in_x} are separated
by at least one zero, the only way that $|m-k| \le 2$
can be satisfied is if $x_k$ is the last symbol of $\xxxi$
whereas $x_m$ is the first symbol of $\xxx_{i+1}$. 
But then the alternating signs rule implemented in \eq{x1}\,--\,\eq{xr}
guarantees that $x_k \ne x_m$. We have thus arrived at a contradiction.
This implies that $x_k$ and $x_m$ (and, hence, also $x_l$) must 
belong to the same sub-block $\xxxi$ of $\xxx$ in \eq{runs_in_x}.

{\bf Case\,2:} 
$x_k, x_l, x_m$ belong to the sub-block $\xxxi$ of length $j$.\\
First suppose that $j \,{\notin}\, \{3,5\}$. Then \eq{x1}\,--\,\eq{xr}
guarantee that $\xxxi = \psi(\one^j)$ or $\xxxi = -\psi(\one^j)$.
For this case, \Lref{psi-lemma} implies that
$x_{k+l-m}$, $x_{k+m-l}$, and $x_{l+m-k}$ also lie within $\xxxi$. This, 
in turn, guarantees that they are all nonzero, 
which is in agreement with the TGP$(2)$ constraint.
We are thus left to deal with the situation where $j = 3$ or $j=5$.
This is precisely where the forbidden blocks in $\cF(2)$ come into
play.
\begin{list}{}
{
\addtolength{\leftmargin}{-2.00ex}
\setlength{\rightmargin}{\leftmargin}
}
\item
\hspace*{2ex}{\bf Case\,2.1:} The sub-block $\xxxi$ is of length $j=5$.\\
The key point is that the binary sequence $(001111100)$ never
occurs as a sub-block of $\yyy$. Hence $\xxxi$ never appears 
in the context $\cdots 00\,\xxx_i\,00 \cdots$. Note that the only
relevant context for the TGP$(2)$ constraint consists of the 
two symbols immediately before $\xxxi$ and the two symbols 
immediately after $\xxxi$. The fact that $(001111100)$ does not
occur in $\yyy$ together with the encoding rules in \eq{psi}\,--\,\eq{xr}
guarantee that $\xxxi$ appears as follows in all of its possible contexts:
\be{5-contexts}
\newcommand{\mydots}{{\cdot}{\cdot}{\cdot}}
\begin{array}{c}
\begin{array}{rr}
({+}{-}{+}{+}{-}00 \mydots  & ({+}{-}{+}{+}{-}0{+} \mydots \\[-0.15ex]
(0{+}{-}{+}{+}{-}00 \mydots & (0{+}{-}{+}{+}{-}0{+} \mydots 
\end{array}
\\[1.50ex]
\begin{array}{c@{\hspace{2ex}}c}
\mydots{+}0{-}{+}{-}{-}{+}00\mydots & \mydots{+}0{-}{+}{-}{-}{+}0{-}\mydots \\[-0.15ex]
\mydots{-}0{+}{-}{+}{+}{-}00\mydots & \mydots{-}0{+}{-}{+}{+}{-}0{+}\mydots \\[-0.15ex]
\hspace{.75ex}\mydots00{+}{-}{-}{+}{-}0{+}\mydots\hspace*{-.75ex} & 
\hspace{.25ex}\mydots00{-}{+}{+}{-}{+}0{-}\mydots\hspace*{-.25ex}
\end{array}
\\[2.50ex]
\begin{array}{r@{\hspace{2ex}}r}
\mydots00{+}{-}{-}{+}{-})  & \mydots00{-}{+}{+}{-}{+}) \\[-0.15ex]
\mydots{-}0{+}{-}{-}{+}{-})  & \mydots{+}0{-}{+}{+}{-}{+}) \\[-0.15ex]
\mydots00{+}{-}{-}{+}{-}0)\hspace*{-1.20ex} & 
\mydots00{-}{+}{+}{-}{+}0)\hspace*{-1.20ex} \\[-0.15ex]
\mydots{-}0{+}{-}{-}{+}{-}0)\hspace*{-1.20ex}  & 
\mydots{+}0{-}{+}{+}{-}{+}0)\hspace*{-1.20ex} 
\end{array}
\end{array}
\ee
where `$($' and `$)$' signify the beginning and the end of the
entire sequence $\xxx\,{=}\,\Psi(\yyy)$, respectively. It is now easy
to verify by direct inspection that each of the $18$ sequences
in \eq{5-contexts} satisfies the TGP$(2)$ constraint.\vspace{.75ex}

\hspace*{2ex}{\bf Case\,2.2:} The sub-block $\xxxi$ is of length $j=3$.\\
Similarly to the previous case, the fact that $(011100)$ and 
$(001110)$ do not occur in $\yyy$ together with the encoding rules
in \eq{psi}\,--\,\eq{xr}
guarantee that $\xxxi$ appears as follows in all of its possible contexts:
\be{3-contexts}
\newcommand{\mydots}{{\cdot}{\cdot}{\cdot}}
\begin{array}{c}
\begin{array}{ccc}
({+}{+}{-}00\mydots & ({+}{+}{-}0{+}\mydots & (0{+}{-}{+}0{-}\mydots
\end{array}
\\[.50ex]
\begin{array}{c@{\hspace{2ex}}c}
\mydots{-}0{+}{-}{+}0{-}\mydots & \mydots{+}0{-}{+}{-}0{+}\mydots 
\end{array}
\\[.50ex]
\begin{array}{lll}
\mydots00{+}{-}{-}) & \mydots{+}0{-}{+}{+}) & \mydots{-}0{+}{-}{+}0)\\
\mydots00{-}{+}{+}) & \mydots{-}0{+}{-}{-}) & \mydots{+}0{-}{+}{-}0).
\end{array}
\end{array}
\ee
Again, it can be verified by direct inspection that each of the $11$ 
sequences in \eq{3-contexts} satisfies the TGP$(2)$ constraint.
\end{list}

\noindent
Since our analysis in Cases 1 and 2 is exhaustive,
this~estab\-lishes property {\bf P2} and completes
the proof of the lemma.
\vspace{1.5ex}
\end{proof}

\Lref{S3n2_sufficiency_lemma} shows that every sequence $\yyy \in \SF(n)$
can be converted to a ternary sequence in $\cT_{3;2}(n)$ 
by assigning $+/-$ labels to the ones in $\yyy$. This implies that 
$\SF(n) \subseteq \cB_{3;2}(n)$, 
by the definition of $\cB_{3;2}(n)$ in \eq{Bqt-def}.
Together with \Lref{S3n2_necessity_lemma}, this completes
the proof of \Tref{S3n2_theorem}. The next corollary 
uses this result 
to determine the capacity of the TGP$(2)$ constraint.

\begin{corollary}
\label{H32_cor}
Let $\rho$ denote the largest-magnitude 
root of the polynomial\, 
$z^{10} - 2z^9 + z^5 - z^4 + 2z^3 - z^2 - 2z + 1$. 
Then the capacity of the TGP\/$(2)$ constraint is given by
\be{cor12}
H_3(2) \,=\, \log_2\rho \,\approx\, 0.96048.
\ee
\end{corollary}

\begin{proof}
We will use the results of Wilf~\cite{Wilf} and of 
Guibas and Odlyzko~\cite{GO}, which provide a much 
more efficient means to compute the capacity of 
a constraint from its set of forbidden blocks than
the standard methods (briefly discussed at the end of this 
subsection).
Let $g_0 = 1$, and for $n \in \Z^+$, define
$$
g_n \,\ \deff\,\
|\cB_{3;2}(n)| 
\, = \, 
\left|\SF(n)\right|.
$$
Further, define the generating function 
$G(z) = \sum_{n=0}^{\infty} g_n z^{-n}$.
Using Theorem\,1 of \cite{GO}, we 
find that $G(z)$ is given by 
$$
G(z)
\ = \
\frac{z(z+1)(z^8-z^7+z^6-z^5+z^4-z^2+2z-1)}
     {z^{10} - 2z^9 + z^5-z^4+2z^3-z^2-2z+1}.
$$
It can be easily verified (using, say, {\sc Matlab} or {\sc Mathemat\-ica})
that the largest-magnitude pole of $G(z)$ is the unique largest-magnitude 
root of its denominator polynomial.
Moreover, this root $\rho$ is real and simple. 
It now follows from the theory of generating functions due
to Wilf~\cite[Chapter\,5]{Wilf} that 
$g_n = \alpha\, \rho^n \bigl(1+o(1)\bigr)$ for some constant $\alpha > 0$. 
Consequently, 
$$
H_3(2) \,=\, \lim_{n \to \infty} \frac{\log_2|\cB_{3;2}(n)|}{n}
       \,=\, \lim_{n \to \infty} \frac{\log_2 g_n}{n} 
       \,=\, \log_2 \rho.
$$
Using the {\sc Mathematica} software package, 
we have found that $\rho \approx 1.94596$, and therefore 
$H_3(2) \approx 0.96048$.
\vspace{1.5ex}
\end{proof}

\looseness=-1
Observe that $H_3(t)$ is \emph{much} larger than $H_2(t)$ 
for $t=1,2$, as can be seen by comparing Table\,\ref{H2t_table}
with \Tref{S3n1_theorem} and \Cref{H32_cor}.
Furthermore, the drop from $H_3(1)$ to $H_3(2)$ is significantly 
smaller than the drop from $H_2(1)$ to $H_2(2)$. 
As mentioned in Section\,\ref{sec:1D}, if this trend 
continues for larger values of $t$,
we can have reasonably efficient codes that, under the
simplifying assumption of that section, 
mitigate the formation of ghost pulses in 
a typical optical communication scenario. 



To conclude our discussion of the TGP(2) constraint, we comment 
upon the design of encoders for converting {arbitrary} binary 
sequences into TGP(2)-constrained ternary sequences. 
The function $\Psi$ constructed in \eq{psi}\,--\,\eq{xr}
provides an explicit method of transforming sequences in 
$\SF(n) = \cB_{3;2}(n)$ into sequences in $\cT_{3;2}(n)$. 
However, this function does not work for \emph{arbitrary} binary 
sequences: if $\yyy {\notin} \SF(n)$, then $\Psi(\yyy)$ is not
necessarily in $\cT_{3;2}(n)$. 
Thus, we still need to design an encoder that converts 
an arbitrary (unconstrained) binary sequence to a sequence 
in the constrained system
$$
\SF
\,\ \deff\,\, \
\smash{\bigcup}_{\hspace*{-2.75ex}\raisebox{-1.50ex}{$\scriptstyle n=1$}}^%
{\hspace*{-2.0ex}\raisebox{1.50ex}{$\scriptstyle \infty$}}
\smash{\SF(n)}. 
$$

The theory of constrained coding provides a standard way 
to design such encoders, which we briefly outline in what 
follows. Let $\cG$ be a {finite, labeled, directed graph}.
We say that $\cG$ is a \emph{presentation of a
constrained system $\cS$} if $\cS$ is the set 
of all sequences obtained by reading the labels of 
all finite paths in $\cG$. A presentation $\cG$ of $\cS$ is 
\emph{deterministic} if at each vertex of $\cG$, 
the outgoing edges are labeled distinctly. 
Given a deterministic presentation of $\cS$ 
along with integers~ $p$ and $q$ such that $p/q$ is 
less than or equal to the capacity
of $\cS$, there is a systematic algorithm~\cite[Section~4]{HCT} 
for designing a rate $p{:}q$ finite-state encoder for $\cS$ along with 
a corresponding decoder. Thus, to construct a finite-state encoder for
our constrained system $\SF \,{=}\: \cB_{3;2}$, all we need to do is 
provide a deterministic presentation for $\SF$.
From this, the desired encoder 
can be generated via the algorithm mentioned above. 

\begin{figure}[ht]
\epsfxsize = 2.75in
\centerline{\epsfbox{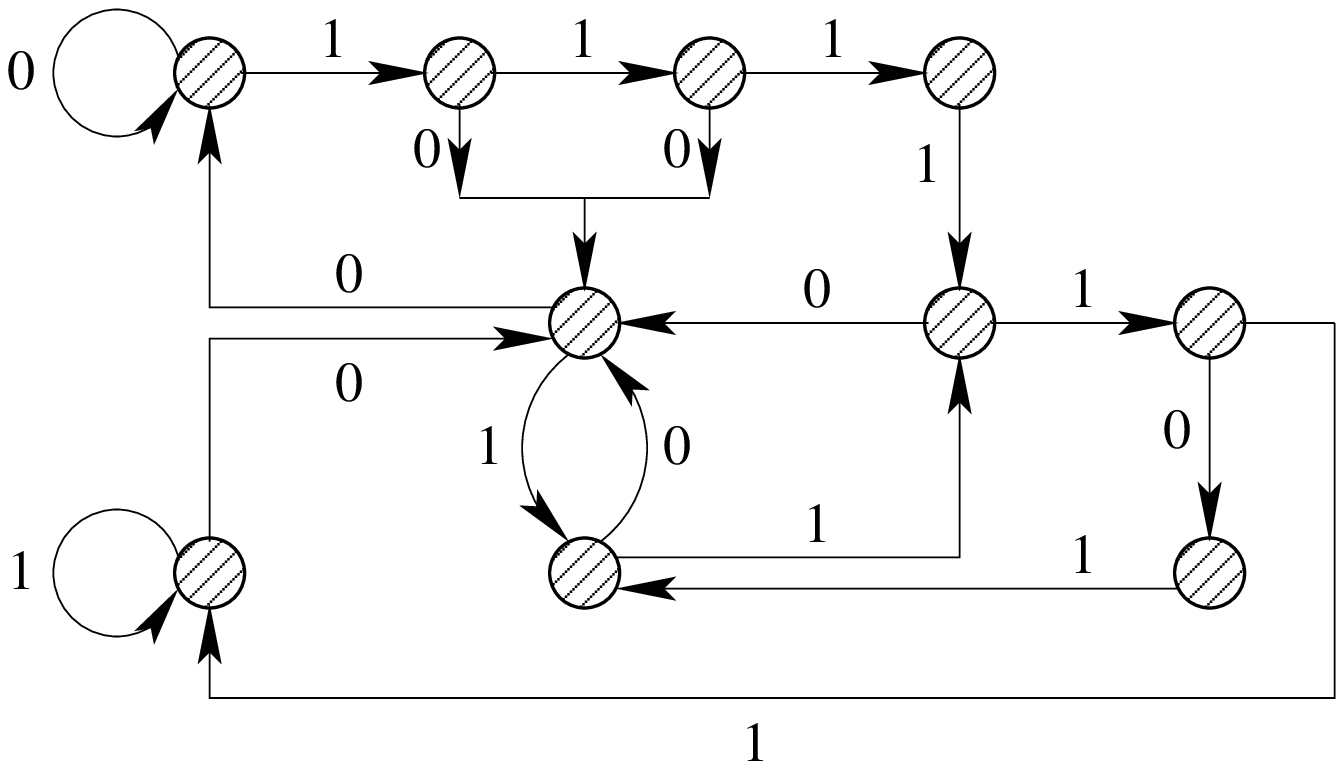}}
\caption{\hspace*{-1.0ex}A deterministic presentation 
of the constrained system $\SF = \cB_{3;2}$}
\label{S32_shannoncover}
\end{figure}

\looseness=-1
It may be verified that the graph in Fig.\,6
is a deterministic presentation of $\SF$. Hence, it 
can be used as the starting point for the design of 
encoders that convert unconstrained binary
sequences to sequences in $\cB_{3;2}$ (and then,
via the function $\Psi$ in \eq{psi}\,--\,\eq{xr}, 
to sequences in $\cT_{3;2}$).
In fact, the graph in Fig.\,6 is the 
\emph{minimal} deterministic presentation
(also known as the \emph{Shannon cover}) of $\cB_{3;2}$, 
in the sense that it has the least number of vertices 
among all deterministic presentations of $\cB_{3;2}$.

While on the subject of deterministic presentations,
let us state the following well-known fact~\cite[Theorem~3.12]{HCT}, 
which will be needed in the next subsection. If $\cG$ is 
a deterministic presentation of a given constrained system $\cS$, 
then the capacity of $\cS$ is 
$\log_2\! \lambda\bigl(A_{\cG}\bigr)$, 
where $\lambda\bigl(A_{\cG}\bigr)$ is the
largest eigenvalue of the adjacency matrix of $\cG$. 
Incidentally, 
this provides an alternative proof of \Cref{H32_cor}, since the 
characteristic polynomial of the adjacency matrix of the graph in 
Fig.\,6 is precisely $z^{10} - 2z^9 + z^5-z^4+2z^3-z^2-2z+1$.

\subsection{The TGP$(t)$ Constraints for $t \ge 3$}
\label{tgpt_section}

\noindent
It is clear that the painstaking analysis presented in the previous 
subsection cannot be easily extended to the TGP$(t)$ constraint for an 
arbitrary $t \in \Z^+$. Instead, we suggest an alternative, systematic 
approach to tackle the general case, which can, in principle, be 
programmed into a computer.

The approach developed in this section has two main disadvantages.
First, instead of computing $H_3(t)$ we end up with a slightly
different quantity
\be{H't}
\hH_3(t) \,\ \deff\,\ \lim_{n\to\infty} \frac{\log_2|\hB(n)|}{n}
\ee
where $\hB(n)$ is the set of all binary sequences
of length~$n$ that can be \emph{extended to a bi-infinite sequence}
without violating the TGP$(t)$ constraint (more precise definition
to follow shortly).
This is not much of a problem, since $\hH_3(t) \le H_3(t)$ for all $t$
and there are good reasons to believe that 
$\hH_3(t) = H_3(t)$ for all $t$ (see the remark below).
The second problem is the computational complexity of the proposed
approach. Unfortunately, this complexity is doubly-exponential in $t$.
In fact, in order to compute $\hH_3(t)$ one needs to construct a graph
with at least $\smash{2^{\Omega(9^t)}}$ vertices. Thus the proposed
approach is not practical even for $t=2$. Nevertheless, we believe
that this approach has conceptual value, and sheds additional light
on the underlying structure of the TGP$(t)$ constraint.

\looseness=-1
The general idea behind our approach is to develop a procedure that, 
given a $t \in \Z^+$, generates a deterministic presentation $\G_{3;t}$
of the constrained system 
\be{hB1}
\hB
\,\ \deff\, \,\
\smash{\bigcup}_{\hspace*{-2.75ex}\raisebox{-1.50ex}{$\scriptstyle n=1$}}^%
{\hspace*{-2.0ex}\raisebox{1.50ex}{$\scriptstyle \infty$}}
\smash{\hB(n)}. 
\ee
In developing our results, it would be much more convenient 
to deal with bi-infinite sequences. This eliminates the 
``edge effects'' present at the beginning and end of 
a finite sequence, which could be quite bothersome 
(for example, much of the effort in describing the 
encoding rule $\Psi$ of the previous subsection 
---\,see\,\eq{x1},\,\eq{xi-mod},\,\eq{xr}\,--- was 
devoted to such edge effects).\vspace{0.50ex}

\looseness=-1
Recall that 
$\cT^*_3$ was defined in 
Section\,\ref{tgp_section} as the set of bi-infinite
ternary sequences satisfying the\kern-.75pt\ TGP 
constraint.\kern-0.50pt\ We extend 
this definition in the natural way to the TGP$(t)$ 
constraint.

\begin{definition}
\label{tbinfinite_def}
A bi-infinite sequence $\xxx \,{=}\, \smash{\{x_j\}_{j \inn \Z}}$ over 
the ternary alphabet $\cA_3 \,{=}\, \{-1,0,1\}$ 
is said to satisfy the TGP\/$(t)$ constraint if 
for all $k,l,m \in \Z$ such that 
$$
\max\bigl\{|k\,{-}\,l|,|l\,{-}\,m|,|m\,{-}\,k|\bigr\} \,\leq\, t \,,
$$
whenever $x_k$, $x_l$, $x_m$ are equal and nonzero, then  
$x_{k+l-m}$ is also nonzero. 
We let $\cT^*_{3;t}$ denote the set of all bi-infinite 
ternary sequences satisfying the TGP\/$(t)$ constraint, 
and let $\cB^*_{3;t} {=}\, \xi\bigl(\cT^*_{3;t}\bigr)$ 
denote the set of all bi-infinite {binary} sequences 
that can be converted to a sequence in $\cT^*_{3;t}$ 
by negating some of their ones. 
\end{definition}

\hspace*{-2.5pt}We now construct a deterministic presentation for 
$\cT^*_{3;t}$. Given
a $t \in \Z^+$, define a finite, labeled, directed 
graph $\hG_{3;t}$, as follows. The set of vertices of 
$\hG_{3;t}$ is the set of all
$$
\xxx 
= 
\bigl(
x_{-t}x_{-t+1} \ldots x_{-1}x_0x_1 \ldots x_{2t-1}x_{2t}
\bigr) \in \cA_3^{3t+1}
$$
that satisfy the following condition: 
for all 
$
k,l,m \in [0,t]
$ 
such that $x_k$,\kern1.5pt$x_l$,\kern1.5pt$x_m$ are equal and 
nonzero, we also have $x_{k+l-m} {\ne}\, 0$.
Note that the position indices $k,l,m$ are restricted 
to the interval $[0,t]$ in the above condition. This
implies that $\hG_{3;t}$ has at least
$3^{2t}$ 
vertices; for example all the 
sequences of the form
$$
\bigl(x_{-t} x_{-t+1} \ldots x_{-1} \, 0 0 \ldots 0 \:
 x_{t+1} x_{t+2} \ldots x_{2t}\bigr)
$$
are vertices of $\hG_{3;t}$. In fact, the order 
(number of vertices) of 
$\hG_{3;t}$ is probably closer to $3^{3t}$ than to $3^{2t}$
(however, when $t$ is small, 
the vertices of $\hG_{3;t}$ can still be enumerated by
exhaustive computer search).
The edges of $\hG_{3;t}$ are defined as follows.
For each pair of vertices
$$
\xxx = \bigl(x_{-t} x_{-t+1} \ldots x_{2t}\bigr)
\hspace{2ex}\text{~and~}\hspace{2ex}
\xxx' = \bigl(\hx_{-t} \hx_{-t+1} \ldots \hx_{2t}\bigr)
$$
where $\xxx$ and $\xxx'$ are not necessarily distinct, 
we draw a single directed edge from $\xxx$ to $\xxx'$ 
if and only if the last $3t$ symbols of $\xxx$ are equal
to the first $3t$ symbols of $\xxx'$, that is if
$$
\bigl(x_{-t+1}x_{-t+2} \ldots x_{2t}\bigr) 
\,=\,
\bigl(\hx_{-t}\hx_{-t+1} \ldots \hx_{2t-1}\bigr).
$$
The label of this directed edge is the symbol $\hx_{2t}$. 
This completes our construction of the graph $\hG_{3;t}$.

Given a finite, labeled, directed graph $\cG$, the 
\emph{sofic shift of $\cG$} is the set of all bi-infinite 
sequences obtained by reading the labels of \emph{bi-infinite} 
paths in $\cG$. One of our main results in this subsection 
is the following theorem.

\begin{theorem}
\label{sofic}
Let $\cX_{3;t}$ denote the sofic shift 
of the graph $\hG_{3;t}$.
 Then, for all $t \in \Z^+$, we have\vspace{-.50ex} 
\be{th13}
\cX_{3;t} 
\, = \, 
\cT^*_{3;t}.
\ee
\end{theorem}

\begin{proof}
\looseness=-1
We first show that $\cT^*_{3;t} \subseteq \cX_{3;t}$.
Consider any element $\xxx \,{=}\, \smash{\{x_j\}_{j \inn \Z}}$ 
of $\cT^*_{3;t}$. For all $j \in \Z$, let $\xxx_j$ denote
the sub-block $(x_{j-t}x_{j-t+1} \ldots x_{j+2t})$
of $\xxx$. Since $\xxx$ 
satisfies the TGP$(t)$ constraint, it follows from our construction 
of $\hG_{3;t}$ that $\xxx_j$ is a vertex of $\hG_{3;t}$ for
all $j \in \Z$. Moreover, since the last $3t$ symbols of
$\xxx_{j-1}$ are obviously equal to the first $3t$ symbols of $\xxx_j$, 
the graph $\hG_{3;t}$ has a unique edge $e_j$ from $\xxx_{j-1}$ to $\xxx_j$,
which is labeled by $x_{j+2t}$. But then, the sequence of such
edges $\smash{\{e_j\}_{j \inn \Z}}$ 
is a path in $\hG_{3;t}$ that generates $\xxx$. It follows
that $\xxx \in \cX_{3;t}$. 

\looseness=-1
In order to establish the inclusion $\cX_{3;t} \subseteq \cT^*_{3;t}$, 
consider any element $\xxx \,{=}\, \smash{\{x_j\}_{j \inn \Z}}$ 
of $\cX_{3;t}$ and let $\smash{\{e_j\}_{j \inn \Z}}$ denote the 
path in $\hG_{3;t}$ that generates $\xxx$. We again let 
$\xxx_j$ denote the sub-block
$(x_{j-t}x_{j-t+1} \ldots x_{j+2t})$
of $\xxx$. 
Then, it follows from our construction of 
$\hG_{3;t}$ that for all $j \in \Z$,
the vertex at which $e_{j+2t}$ terminates 
must be the sequence $\xxx_j$. 
Therefore,
$\xxx_j$ is a vertex in $\hG_{3;t}$
for all $j \in \Z$. 
Now, suppose we have $k,l,m \in\! \supp(\xxx)$ 
such that 
$\max\{|k{-}l|,|l{-}m|,|m{-}k|\} \leq t$ and $x_k = x_l = x_m$.
In order to prove that $\xxx \in \cT^*_{3;t}$, 
we must show that $x_{k+l-m} {\neq}\, 0$. 
Since $k,l,m$ are all within a distance of $t$ of each other,
there exists a $j \in \Z$ such that $k,l,m \in [j,j\,{+}\,t]$. 
Observe that for any $k,l,m \in [j,j+t]$, the integer
$k\,{+}\,l\,{-}\,m$ lies in $[j\,{-}\,t,j\,{+}\,2t]$. But now, 
since $\xxx_j$ is a vertex of $\hG_{3;t}$, it follows from 
our definition of the vertex set of $\hG_{3;t}$ that $x_{k+l-m} \neq 0$.
Thus $\xxx \in \cT^*_{3;t}$, which shows that 
$\cX_{3;t} \subseteq \cT^*_{3;t}$ and completes the proof.
\vspace{1.5ex}
\end{proof}

We now define $\cT'_{3;t}$ as the set of all finite-length
sequences that are sub-blocks of some sequence in $\cT^*_{3;t}$.
Stated another way, $\cT'_{3;t}$ is a subset of the set $\cT_{3;t}$
defined in Section\,\ref{def_section}, consisting of all finite-length
sequences that a) satisfy the TGP$(t)$ constraint and b) can 
be extended to a bi-infinite sequence that satisfies the 
TGP$(t)$ constraint. It is possible that some finite-length
sequences in $\cT_{3;t}$ \emph{cannot} be extended in this way, 
in which case $\cT'_{3;t}$ is strictly smaller than $\cT_{3;t}$.

\begin{corollary}
\label{Cor14}
Let $X_{3;t}$ denote the constrained system of the graph $\hG_{3;t}$.
Then, for all $t \in \Z^+$, we have
\be{cor14}
X_{3;t} 
\, = \,
\cT'_{3;t}.
\ee
Moreover, the graph $\hG_{3;t}$ is a deterministic
presentation of its constrained system $X_{3;t} = \cT'_{3;t}$.
\end{corollary}

\begin{proof}
It should be obvious from our construction of $\hG_{3;t}$
that outgoing edges at each vertex of $\hG_{3;t}$ are
labeled distinctly.
Hence $\hG_{3;t}$ is a deterministic
presentation of its constrained system. Furthermore, it 
is well known (and obvious) that \eq{th13} implies \eq{cor14}.
In the terminology of symbolic dynamics, the sets $X_{3;t}$
and $\cT'_{3;t}$ are precisely the \emph{languages} of the
sofic shifts $\cX_{3;t}$ and $\cT^*_{3;t}$. Since the shifts
are equal (by \Tref{sofic}), their languages must be also equal.
\vspace{1.25ex}
\end{proof}

\Cref{Cor14} implies that we can find 
the capacity of $\cT'_{3;t}$ from the largest eigenvalue of the 
adjacency matrix of $\hG_{3;t}$. 
However, we are not interested in $\cT'_{3;t}$, but rather in the set
\be{hB2}
\hB
\, = \, 
\xi\bigr(\cT'_{3;t}\bigr)
\, = \,
\bigl\{
\, \xi(\xxx) \,:\, \xxx \in \cT'_{3;t} \,
\bigr\}.
\ee
Letting $\hB(n)$ denote the number of sequences of length~$n$
in $\hB$, we get the expression \eq{H't} for the 
capacity $H'_3(t)$.
\vspace{1.5ex}

\noindent
{\bf Remark.}
Here is a heuristic argument in support 
of our claim that $\hH_3(t)$ is likely 
to be equal to $H_3(t)$. The difference between 
$\hH_3(t)$ and $H_3(t)$ stems 
from the difference between the sets
$\cT'_{3;t}$ and $\cT_{3;t}$. 
It is well known~\cite{Immink2,LM,HCT} that the 
capacity of a language is equal to the
entropy of the underlying shift. Thus,
instead of looking at $\cT'_{3;t}$, 
we might as well look at the underlying
sofic shift $\cX_{3;t} \,{=}\, \cT^*_{3;t}$.
The TGP$(t)$ constraint defining $\cT_{3;t}$ 
is a finite restriction 
of the TGP$(t)$ constraint defining $\cT^*_{3;t}$.
Furthermore, the TGP$(t)$ constraint is \emph{local},
in the sense that it is defined through a finite 
window of length~$t$. 

\looseness=-1
Now, it is generally observed in the literature~\cite{LM} 
that if a constrained system 
$\cS$ is obtained via a finite restriction of 
a local constraint that defines a sofic shift $\cX$, 
then the capacity of $\cS$ equals the entropy of $\cX$. 
Of course, this is clearly true whenever any finite sequence in $\cS$ can 
be extended to a bi-infinite sequence in $\cX$. 
However, ``edge effects'' sometimes make it impossible 
to extend certain sequences in $\cS$ without
violating the constraint. But, in the case of 
a local constraint, these edge effects are usually not strong
enough to affect a significant proportion of the sequences in $\cS$,
so that the capacity of $\cS$ is still equal to the entropy of $\cX$.  
This is not always true, but the exceptions to this rule tend 
to be pathological. 

It may be possible to prove rigorously that 
$H'_3(t) = H_3(t)$, but such a proof would have
to deal in detail with the ``edge effects'' and
is likely to be too tedious to be worth the effort. 
\vspace{1.5ex}

The remaining problem is to construct a deterministic presentation 
for the set $\hB$ in \eq{hB1} and \eq{hB2}.
Given the graph $\hG_{3;t}$, constructing a presentation
for $\hB$ is easy: simply apply $\xi(\cdot)$ to all the 
labels in $\hG_{3;t}$. Specifically, let $\hG'_{3;t}$
denote the graph obtained from $\hG_{3;t}$ by replacing 
the labels of all the edges 
with their absolute values. Then it is obvious from \eq{hB2}
and Corollary\,\ref{Cor14} that the graph $\hG'_{3;t}$ is a presentation
for $\hB$.

Note, however, that although $\hG_{3;t}$ is a deterministic
presentation of $\cT'_{3;t}$, the graph $\hG'_{3;t}$ is not 
necessarily a \emph{deterministic}  presentation of $\hB$.
Indeed, there may be two edges emanating from the same vertex $\xxx$
in $\hG_{3;t}$, 
one labeled with ${+}$ and the other with ${-}$, 
whose labels in $\hG'_{3;t}$ would both be $1$.
Fortunately, there is a well-known procedure that, 
given an arbitrary presentation of a constrained system, 
constructs a deterministic presentation for it. This procedure
is called the \emph{subset construction method}; it is described in 
detail in \cite[Section\,2.2.1]{HCT} and in \cite[Theorem\,3.3.2]{LM}. 
Applying the subset construction method to the graph $\hG'_{3;t}$, 
we finally obtain a deterministic presentation $\cH_{3;t}$ for 
the set $\hB$. Given this presentation, we can compute the capacity
$\hH_3(t)$ and construct encoders into $\hB$, 
as described in 
the previous subsection.

We can now summarize the entire procedure
for computing the capacity $\hH_3(t)$, as 
follows:\vspace{-1.250ex}
\begin{itemize}
\item[]
\item[\fbox{{\bf 1}}]
Construct the graphs $\hG_{3;t}$ and $\hG'_{3;t}$ as described 
above, and let $\hB$ be the constrained system presented by $\hG'_{3;t}$.
\vspace{.75ex}

\item[\fbox{{\bf 2}}]
Apply the subset construction method to $\hG'_{3;t}$ in order
to obtain a deterministic presentation $\cH_{3;t}$ for $\hB$. 
\vspace{.75ex}

\item[\fbox{{\bf 3}}]
Construct the adjacency matrix $A_{3;t}$ of $\cH_{3;t}$, 
and compute its largest eigenvalue $\lambda = \lambda(A_{3;t})$. 
Set $\hH_3(t) = \log_2\!\lambda$.
\vspace{1.250ex}
\end{itemize}
Of course, in theory, $\cH_{3;t}$ can also be used to construct 
finite-state encoders for converting unconstrained binary sequences 
to sequences in $\hB \subset \cB_{3;t}$, as explained in 
Section~\ref{tgp2_section}.
In turn, the graphs $\hG_{3;t}$ and $\hG'_{3;t}$ provide 
a method for transforming a binary sequence $\yyy \in \hB$
into a ternary sequence $\xxx$, with $\xi(\xxx) = \yyy$, 
that satisfies the TGP$(t)$ constraint.
For each given $\yyy \in \hB$, 
there is a path in $\hG'_{3;t}$ whose 
label sequence is $\yyy$. We may then take $\xxx$ to be the sequence 
of labels along the \emph{same path} in $\hG_{3;t}$. The practicality 
of this method depends on the existence of a systematic procedure for
finding a path in $\hG'_{3;t}$ that generates $\yyy$. 
Of course, it also depends on the order 
of the graphs $\cH_{3;t}$, $\hG'_{3;t}$, and $\hG_{3;t}$.

We have already observed that the order of $\hG_{3;t}$ and $\hG'_{3;t}$
is exponential in $t$. However, since we are interested primarily
in small values of $t$, such exponential growth could still be 
tolerated. The main computational problem is with the subset
construction method at Step\,2 above. The subset construction 
technique, when applied to a graph with $n$ vertices, 
produces a graph with $2^n - 1$ vertices. As a result, 
the graph $\cH_{3;t}$ constructed in Step\,2 
has at least $\smash{2^{9^t}}$ vertices. In fact, this is likely
to be a vast underestimate of the order of $\cH_{3;t}$.

\section{Summary}
\vspace{0.5ex}
\label{summary}

\noindent 
We have defined and analyzed a number of ``ghost-pulse'' 
constraints that can be used to design coding schemes 
which mitigate 
the formation of ghost pulses in the optical fiber channel. 
We show that coding schemes based upon sequences that 
satisfy the \emph{binary} ghost-pulse (BGP) constraint must 
necessarily have poor rates, since the capacity of this constraint 
is zero. Sequences satisfying a more relaxed constraint, which we 
call the BGP($t$) constraint, are more suitable for use as codes;
however, the rate of such codes is still too low for practical 
applications.
%
A more promising approach is to use the phase-modulation
idea, which leads to \emph{ternary} constraints. Thus we
study the ternary ghost-pulse (TGP) and TGP$(t)$ constraints. 
We leave the analysis of the TGP constraint with unbounded memory
as an open problem, conjecturing that it has zero capacity. 
But we do provide a detailed analysis of the TGP$(1)$ and TGP$(2)$ 
constraints. Our analysis suggests that coding schemes using 
TGP$(t)$-constrained sequences can achieve much higher rates
than those using BGP$(t)$-constrained sequences. 
We are therefore led to believe 
that TGP$(t)$ constraints yield reasonably efficient schemes for
mitigating the ghost-pulse problem. We also discuss the design of 
encoders and decoders for coding schemes involving 
the BGP, the BGP$(t)$, and the TGP$(t)$ constraints. While the 
procedures we suggest for coding into the BGP$(t)$, TGP$(1)$,
and TGP$(2)$ constraints can be implemented in practice, the 
corresponding design procedure for the general TGP$(t)$ constraint 
with $t \ge 3$ is too computationally intensive to be 
implementable in its present form.


\section*{Acknowledgment} 

\noindent
We are indebted 
to Nikola Alic, Shaya Fainman,
and George Papen for their assistance in helping us 
understand the physics underlying the ghost-pulse effect.

\vspace{.75ex}


\end{document}